\documentclass[useAMS,usenatbib]{mn2e}
\usepackage{graphicx,fleqn,natbib,amssymb}
\DeclareGraphicsExtensions{.eps,.cps,.ps,.eps.gz,.ps.gz,.eps.Z}
\usepackage{subfigure}
\usepackage{multirow}
\usepackage{captcont}
\usepackage{epsfig}
\usepackage{rotating}

\title[Comparison of the UVOT and XRT GRB Light Curves.]{A Statistical Comparison of the Optical/UV and X-ray Afterglows of Gamma-Ray Bursts using the {\it Swift} Ultra-violet Optical and X-ray Telescopes.}

\author[Oates et al.]{S. R. Oates$^{1}$, M. J. Page$^{1}$, P. Schady$^{1}$, M. De Pasquale$^{1}$, P. A. Evans$^{2}$, K. L. Page$^{2}$, 
\newauthor M. M. Chester$^{3}$, P. A. Curran$^{1}$, T. S. Koch$^{3}$, N. P. M. Kuin$^{1}$,  P. W. A. Roming$^{3}$, 
\newauthor M. H. Siegel$^{3}$, S. Zane$^{1}$ and  J. A. Nousek$^{3}$\\
$^{1}$ Mullard Space Science Laboratory, University College London, Holmbury St. Mary, Dorking Surrey, RH5 6NT, UK; sro@mssl.ucl.ac.uk \\
$^{2}$ X-ray and Observational Astronomy Group,  Department of Physics and Astronomy, University of Leicester, LE1 7RH, UK\\
$^{3}$ Department of Astronomy and Astrophysics, Pennsylvania State University, 104 Davey Laboratory, University Park, PA 16802} 

\begin{document}

\date{Accepted...Received...}

\maketitle

\label{firstpage}

\begin{abstract} 
We present the systematic analysis of the UVOT and XRT light curves for a sample of 26 {\it Swift} Gamma-Ray Bursts (GRBs). 
By comparing the optical/UV and X-ray light curves, we found that they are remarkably different during the 
first 500s after the BAT trigger, while they become more similar during the middle phase of the afterglow, 
i.e. between 2000s and 20000s.

If we take literally the average properties of the sample, we find that the mean temporal indices 
observed in the optical/UV and X-rays after 500s are consistent with a forward-shock scenario, under the 
assumptions that electrons are in the slow cooling regime, the external medium is of constant density 
and the synchrotron cooling frequency is situated between the optical/UV and X-ray observing bands. While 
this scenario describes well the averaged observed properties, some individual GRB afterglows require 
different or additional assumptions, such as the presence of late energy injection. 

We show that a chromatic break (a break in the X-ray light curve that is not seen in the optical) is present in the afterglows 
of 3 GRBs and demonstrate evidence for chromatic breaks in a further 4 GRBs. The average properties of these breaks 
cannot be explained in terms of the passage of the synchrotron cooling frequency through the observed bands, 
nor a simple change in the external density. It is difficult to reconcile chromatic breaks in terms of a single 
component outflow and instead, more complex jet structure or additional emission components are required.

\end{abstract}


\section{Introduction}
\label{intro}
Gamma-ray bursts (GRBs) are intense flashes of gamma-rays that can last from as little as a few milliseconds up to a 
few thousand seconds after the trigger. The duration and spectral hardness distributions are found to be bimodal, 
leading to a division of GRBs into two classes: short-hard GRBs ($<2$s) and long-soft GRBs ($>2$s) \citep{kou93}. 
The prompt gamma-ray emission is expected to be followed by an afterglow. The afterglow is most commonly seen in 
the X-rays, but is also observed in the optical/UV and, less commonly, down to radio wavelengths. The duration of 
the afterglow in the X-ray and optical/UV band varies considerably from GRB to GRB, and it has been observed to 
last for as little as a few hours up to a few months after the trigger. In the radio, the afterglow emission may 
be detected up to several years after the prompt gamma-ray emission. The afterglow from a short GRB tends to be 
fainter and short-lived in comparison with the long GRBs. For this reason, only long-GRBs fall into the \cite{oates09} 
selection criteria. This paper is the successor to \cite{oates09} and therefore uses the same sample of 26 GRBs.

Due to the unpredictability and rapid fading of these cosmic explosions, crucial clues onto their nature, their possible  
progenitors and their environments could only be obtained through deep and continuous observations of the afterglow. A rapid response 
satellite, {\it Swift}, which was launched in November 2004, was specifically designed to observe these events. {\it Swift} houses 
three instruments designed to capture the gamma-ray, X-ray and optical/UV emission. The Burst Alert Telescope (BAT; \citealt{bar05}) 
detects the prompt gamma-ray emission. The X-ray Telescope (XRT; \citealt{bur05}) and the Ultra-violet/Optical Telescope (UVOT; \citealt{roming}) 
observe the afterglow. The energy ranges of the BAT and the XRT instruments are 15~keV~-~350~keV and 0.2~keV~-~10~keV, respectively, 
and the wavelength range of the UVOT is 1600\AA-8000\AA. The co-alignment of the XRT and UVOT instruments is ideal for observing 
GRB afterglows because observations of the X-ray and optical/UV afterglow are performed simultaneously.

One of the early results that emerged from the first 27 X-ray afterglows collected by {\it Swift} is the existence of a 
``canonical'' X-ray light curve, which typically comprises of 4 segments \citep{nousek}. Here and throughout the paper we 
will use the flux convention $ F\,\propto\,t^{\alpha}\,\nu^{\beta}$ with $\alpha$ and $\beta$ being the temporal and 
spectral indices respectively. With this notation, the canonical X-ray light curve can be described as an initial steep 
decay segment ($-5<\alpha_{\rm X}<-3$) transitioning to a shallow decay phase \citep[$-1.0<\alpha_{\rm X}<0.0$;][]{liang07}, then 
followed by a slightly steeper decay ($-1.5<\alpha_{\rm X}<-1.0$), which finally breaks again at later times. The last segment 
is usually identified as a post jet-break decay \citep{zhang06}. However, the application of this model to all GRBs has 
recently been questioned by \cite{eva09} with a larger sample of 327 GRBs, 162 of which are considered by \cite{eva09} 
to be well-sampled. This paper found that the ``canonical'' behaviour accounts for only $\sim42\%$ of XRT afterglows.

A statistical study of the UVOT light curves has recently shown that, although there are some similarities between the optical/UV and X-ray bands, 
in general the optical/UV afterglow does not behave in the same way as the X-ray one \citep{oates09}. In particular, the optical/UV 
light curves can either decay from the beginning of the observations or exhibit an initial rise and then a decay phase. In both 
cases, the decay segment is usually well fitted by a power-law, although a small number of GRBs require a broken or a doubly 
broken power-law. Moreover, by systematically comparing the optical/UV light curves with the XRT canonical model, \cite{oates09} 
found that among the four segments of the XRT canonical model the shallow decay segment has the most similar range of temporal 
indices to the optical/UV light curves. The temporal indices of the other segments of the XRT canonical light curve are
steeper than the temporal indices of the optical/UV light curves.

In this paper, we present a statistical cross comparison of the XRT and UVOT light curves for a sample of 26 GRBs presented in \cite{oates09}. 
Table \ref{GRBs} lists these GRBs and their respective redshifts. The paper is organized as followed. In \S~\ref{reduction} we 
describe the data reduction and analysis. The main results are presented in \S~\ref{results}. Discussion and conclusions follow 
in \S~\ref{discussion}, \ref{conclusions}, respectively. All uncertainties throughout this paper are quoted at 1$\sigma$.

\section{Data Reduction and Analysis}
\label{reduction}
The 0.3~keV~-~10~keV X-ray light curves were obtained from the GRB light curve repository at the UK {\it Swift} Science Data Center 
\citep{evans07,eva09}. In order to directly compare the behaviour of the UVOT and XRT light curves, we required the bins 
in the XRT light curve to be small, allowing us to rebin the light curve so that the X-ray bins have the same start and 
end times as the corresponding bins in the UVOT light curve. In order to be able to use Gaussian statistics for error 
propagation (when performing background subtraction and corrections due to pile-up and removal of bad columns), the 
minimum binning provided by the XRT repository is 15 counts per bin. We set the binning to be a minimum of 15 counts 
per bin for both the windowed timing and photon counting modes and switched the dynamic binning option off. For some 
of the repository light curves the last data point has a detection of $<3\sigma$. These points are provided by the 
repository as an upper limit and are excluded from further analysis.

The optical/UV light curves were taken from \cite{oates09} (see Section 3.1 of that paper for a detailed description of the 
construction of the UVOT light curves). These light curves are normalized to the $v$ filter and grouped with a binsize 
of $\Delta t/t=0.2$. The X-ray data were then binned so that the X-ray bins had the same time ranges as the UVOT light curve bins. 
The binned X-ray and $v$-band count rate light curves for each GRB can be seen in the top pane of each panel in Fig. \ref{light curves}.  

In \cite{oates09}, the start time of each UVOT light curve was taken to be the start time of the gamma-ray emission rather than 
the BAT trigger time. The start time of the gamma-ray emission we take to be the start time of the $T_{90}$ parameter. This parameter 
corresponds to the time in which 90\% of the counts in the 15~keV~-~350~keV band arrive at the detector \citep{sak07} and is 
determined from the gamma-ray event data for each GRB, by the BAT processing script. The results of the processing are publicly 
available and are provided for each trigger at http://gcn.gsfc.nasa.gov/swift\_gnd\_ana.html. Therefore, to have consistent start 
times, the XRT light curves were adjusted to have the same start times as the UVOT light curves.

We then applied three different techniques to the optical/UV and X-ray light curves to determine how their behaviour compares 
over the course of {\it Swift} observations; these techniques are described in Sections 2.1 to 2.3. To avoid having hardness 
ratios with errors larger than $\pm1$ and to avoid taking the logarithm of negative numbers when determining the root mean 
square deviation we only use the binned data points with a signal to noise ratio $>1$ for these two methods. When determining 
the temporal indices we used all the available data.

\subsection{Optical/UV to X-ray Hardness Ratio}
To determine how the count rates in the optical/UV and X-ray light curves vary with respect to each other, we calculated the 
hardness ratio of the optical/UV and X-ray count rates. We define the hardness ratio $HR$ to be 
\begin {equation}HR=(C_{\rm X}-C_{\rm O})/(C_{\rm X}+C_{\rm O})\end{equation} where $C_{\rm O}$\, is the $v$ band count rate and $C_{\rm X}$\, is the X-ray count rate. 
A hardness ratio equal to -1 indicates that the optical/UV flux is dominant, whereas a $HR=1$ indicates that the X-ray 
flux is dominant. The X-ray and optical/UV light curves have comparable count rates which allows hardness ratios to 
be computed without significant portions of the hardness ratios being saturated. However, the hardness ratios can only 
provide information on the relative spectral change, which may be due to the passage of a synchrotron spectral frequency, 
differences in the emission mechanisms or differences in the emission geometry. The hardness ratios for each GRB can be 
seen in the middle pane of each panel in Fig. \ref{light curves}.

\begin{figure*}
\begin{center}
\includegraphics[angle=-90,scale=0.34]{GRB050319_XRT_UVOT.txt_XOratio.cps}
\includegraphics[angle=-90,scale=0.34]{GRB050525_XRT_UVOT.txt_XOratio.cps}
\includegraphics[angle=-90,scale=0.34]{GRB050712_XRT_UVOT.txt_XOratio.cps}
\includegraphics[angle=-90,scale=0.34]{GRB050726_XRT_UVOT.txt_XOratio.cps}
\includegraphics[angle=-90,scale=0.34]{GRB050730_XRT_UVOT.txt_XOratio.cps}
\includegraphics[angle=-90,scale=0.34]{GRB050801_XRT_UVOT.txt_XOratio.cps}
\end{center}
\captcont[The 26 GRB X-ray and optical/UV afterglows]{The 26 GRB X-ray and optical/UV afterglows. The dotted lines divide the light curves in to the epochs 
(a) to (d), which are $<500$s, 500s-2000s, 2000s-20000s and $>20000$s, respectively. The top pane of each panel shows the X-ray and optical/UV (equivalent to $v$-band) 
light curves. The X-ray light curves (blue triangles) have been binned to have the same bin sizes as the optical/UV data (red circles). 
The middle pane of each panel shows the X-ray to optical/UV hardness ratio, given by Hardness Ratio=($C_{\rm X}-C_{\rm O})/(C_{\rm O}+C_{\rm X}$) 
where $C_{\rm O}$\, is the $v$ band count rate and $C_{\rm X}$\, is the X-ray count rate. The bottom pane of each panel shows 
the root mean square deviation of the logarithmic X-ray light curves relative to the logarithmic, normalized optical/UV light curves in 
a time window 1 dex wide. The window was shifted in steps of 0.15 in log time and the rms deviation was calculated for each window.}
\label{light curves}
\end{figure*}

\begin{figure*}
\begin{center}
\includegraphics[angle=-90,scale=0.34]{GRB050802_XRT_UVOT.txt_XOratio.cps}
\includegraphics[angle=-90,scale=0.34]{GRB050922c_XRT_UVOT.txt_XOratio.cps}
\includegraphics[angle=-90,scale=0.34]{GRB051109a_XRT_UVOT.txt_XOratio.cps}
\includegraphics[angle=-90,scale=0.34]{GRB060206_XRT_UVOT.txt_XOratio.cps}
\includegraphics[angle=-90,scale=0.34]{GRB060223a_XRT_UVOT.txt_XOratio.cps}
\includegraphics[angle=-90,scale=0.34]{GRB060418_XRT_UVOT.txt_XOratio.cps}
\end{center}
\captcont{Continued.}
\label{light curves2}
\end{figure*}

\begin{figure*}
\begin{center}
\includegraphics[angle=-90,scale=0.34]{GRB060512_XRT_UVOT.txt_XOratio.cps}
\includegraphics[angle=-90,scale=0.34]{GRB060526_XRT_UVOT.txt_XOratio.cps}
\includegraphics[angle=-90,scale=0.34]{GRB060605_XRT_UVOT.txt_XOratio.cps}
\includegraphics[angle=-90,scale=0.34]{GRB060607a_XRT_UVOT.txt_XOratio.cps}
\includegraphics[angle=-90,scale=0.34]{GRB060708_XRT_UVOT.txt_XOratio.cps}
\includegraphics[angle=-90,scale=0.34]{GRB060804_XRT_UVOT.txt_XOratio.cps}
\end{center}
\captcont{Continued.}
\label{light curves3}
\end{figure*}

\begin{figure*}
\begin{center}
\includegraphics[angle=-90,scale=0.34]{GRB060908_XRT_UVOT.txt_XOratio.cps}
\includegraphics[angle=-90,scale=0.34]{GRB060912_XRT_UVOT.txt_XOratio.cps}
\includegraphics[angle=-90,scale=0.34]{GRB061007_XRT_UVOT.txt_XOratio.cps}
\includegraphics[angle=-90,scale=0.34]{GRB061021_XRT_UVOT.txt_XOratio.cps}
\includegraphics[angle=-90,scale=0.34]{GRB061121_XRT_UVOT.txt_XOratio.cps}
\includegraphics[angle=-90,scale=0.34]{GRB070318_XRT_UVOT.txt_XOratio.cps}
\end{center}
\captcont{Continued.}
\label{light curves4}
\end{figure*}

\begin{figure*}
\begin{center}
\includegraphics[angle=-90,scale=0.34]{GRB070420_XRT_UVOT.txt_XOratio.cps}
\includegraphics[angle=-90,scale=0.34]{GRB070529_XRT_UVOT.txt_XOratio.cps}
\end{center}
\caption{Continued.}
\label{light curves5}
\end{figure*}

\subsection{Root mean square deviation}
To determine how closely the data points in the optical/UV and X-ray light curves track each other during a given epoch, 
we determined the root mean square (RMS) of the difference between the logarithmic normalized optical/UV and X-ray 
light curves for multiple epochs such that:\begin{equation}RMS=\sqrt{\frac{\sum{(\log{C_{\rm O}}-\log{C_{\rm X})^2}}}{N}}\label{RMS_eqn}\end{equation} where $N$ is the number of data points. 
For each GRB, the root mean square deviation was calculated using a time window 1 dex (a factor of 10) wide shifted in steps of 0.15 in log time, 
starting from 10s until the end of the observations. The section of X-ray light curve within each window was normalized 
to the corresponding section of optical/UV light curve. This was done by adding a constant term to the logarithmic X-ray light 
curve that minimized the $\chi^2$ between the logarithmic optical/UV and logarithmic X-ray light curves. RMS deviation values 
close to zero indicate that the optical/UV and X-ray light curves behave the same, values larger than zero indicate that the 
light curves do not track each other precisely. 

The starting time of 10s and the movement of the window by 0.15 in log time ensures we are performing the analysis systematically 
and that we can directly compare values of the RMS deviation between two or more GRBs since the RMS deviations have been determined 
from data in the same time ranges. The size of the window implies that the value of the RMS deviation will only change when there 
is large scale temporal change in the light curve for instance flaring behaviour or changes in the temporal index of the X-ray 
and/or optical/UV light curves. There are RMS deviation values which were determined across periods when an observing gap occurs, 
typically between 1000s-3000s, because the window over which we determine the RMS deviation is larger than the observing gap.

The errors were determined using: \begin {equation}RMS_{error}=\sqrt{\frac{\sum{e_{\rm X}^2+e_{\rm O}^2}}{N}}\end{equation} 
Since converting the count rate into logarithmic count rate causes the error bars to be asymmetric $e_{\rm X}$ is taken to be the average 
positive and negative errors of $\log{C_{\rm X}}$ and $e_{\rm O}$ is taken as the average positive and negative error of $\log{C_{\rm O}}$. 
The RMS deviation and error is shown in the bottom pane of each panel in Fig. \ref{light curves}.

\begin{figure}

\vspace{0.2cm}
\includegraphics[angle=-90,scale=0.27]{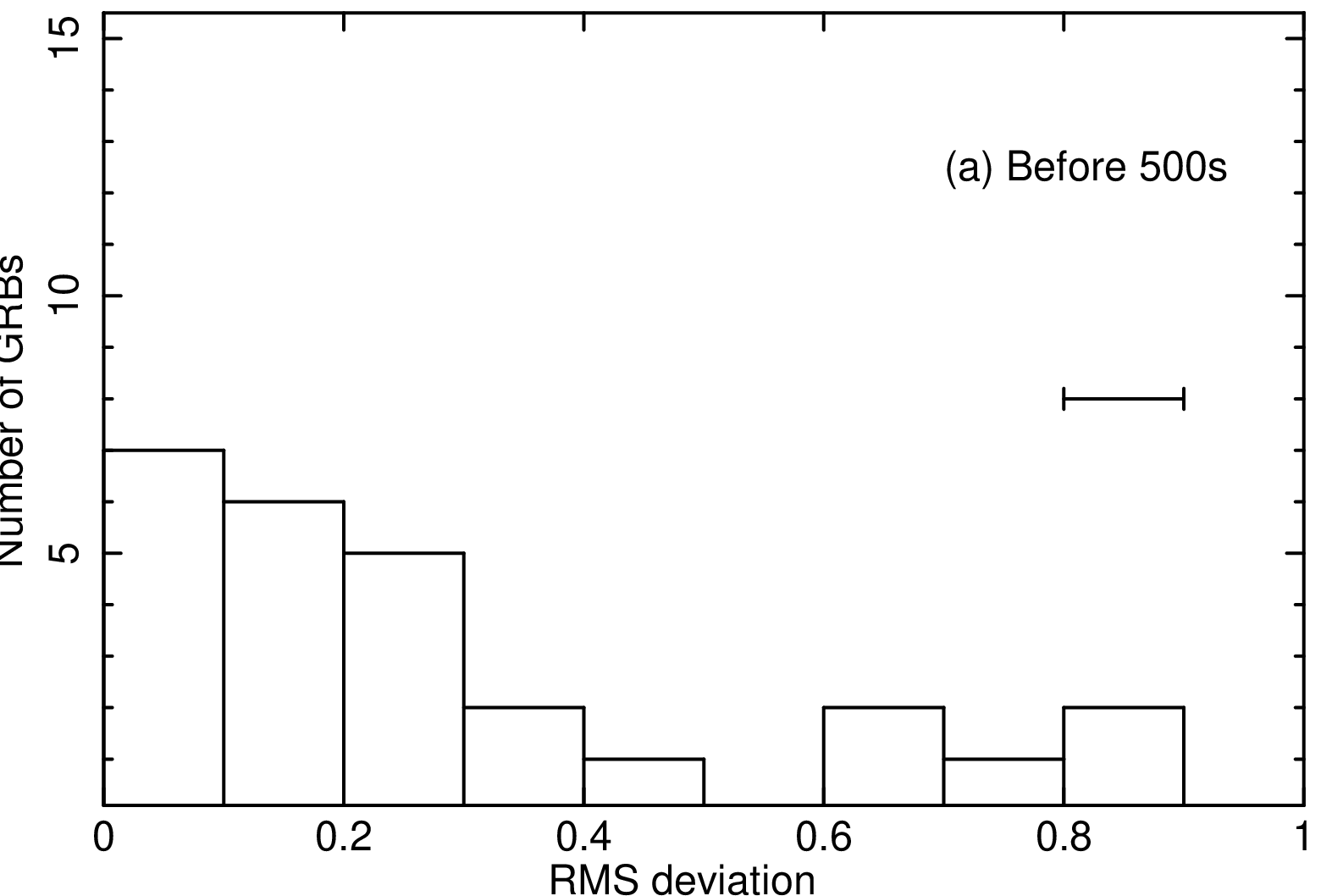}
\vspace{0.2cm}
\includegraphics[angle=-90,scale=0.27]{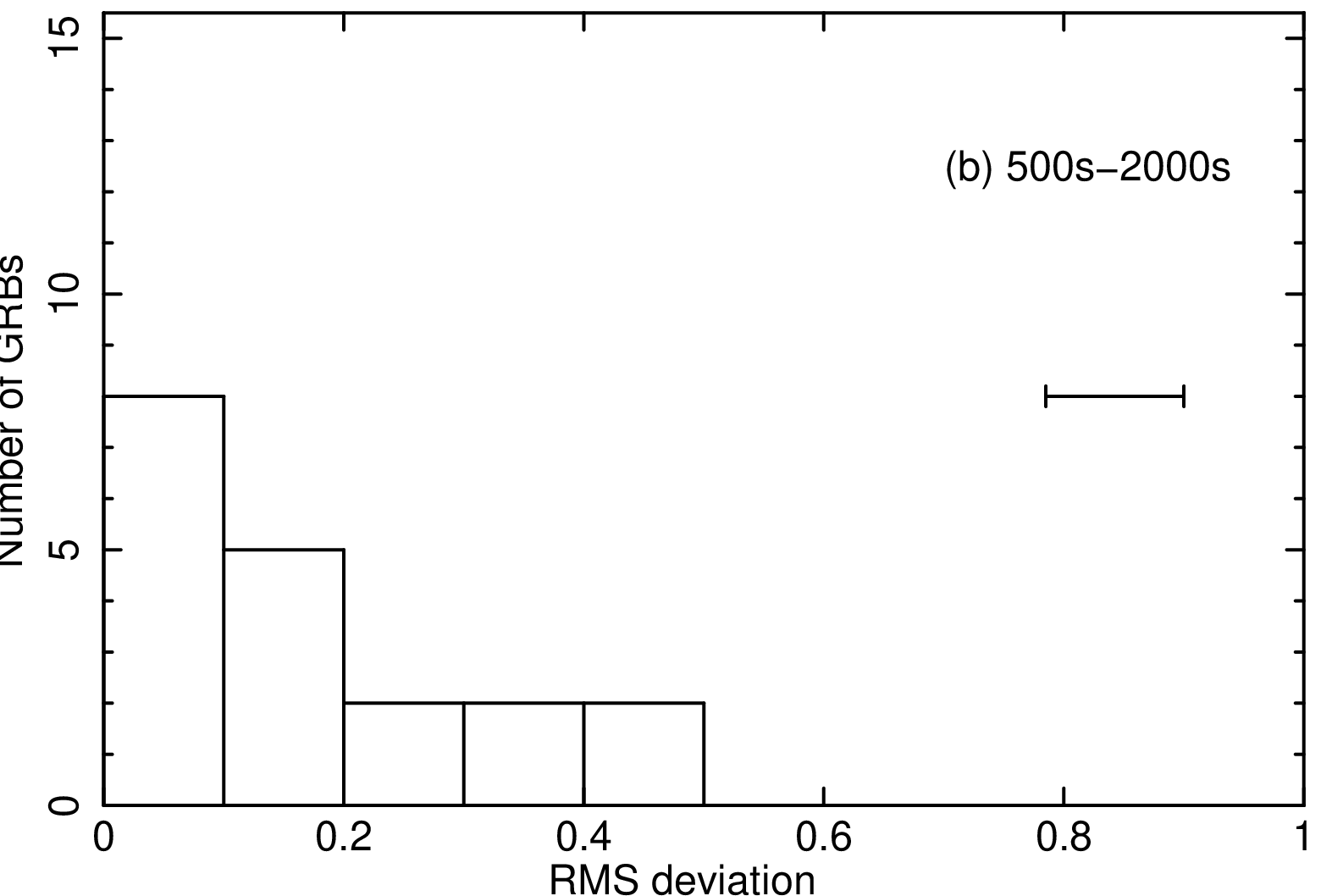}
\vspace{0.2cm}
\includegraphics[angle=-90,scale=0.27]{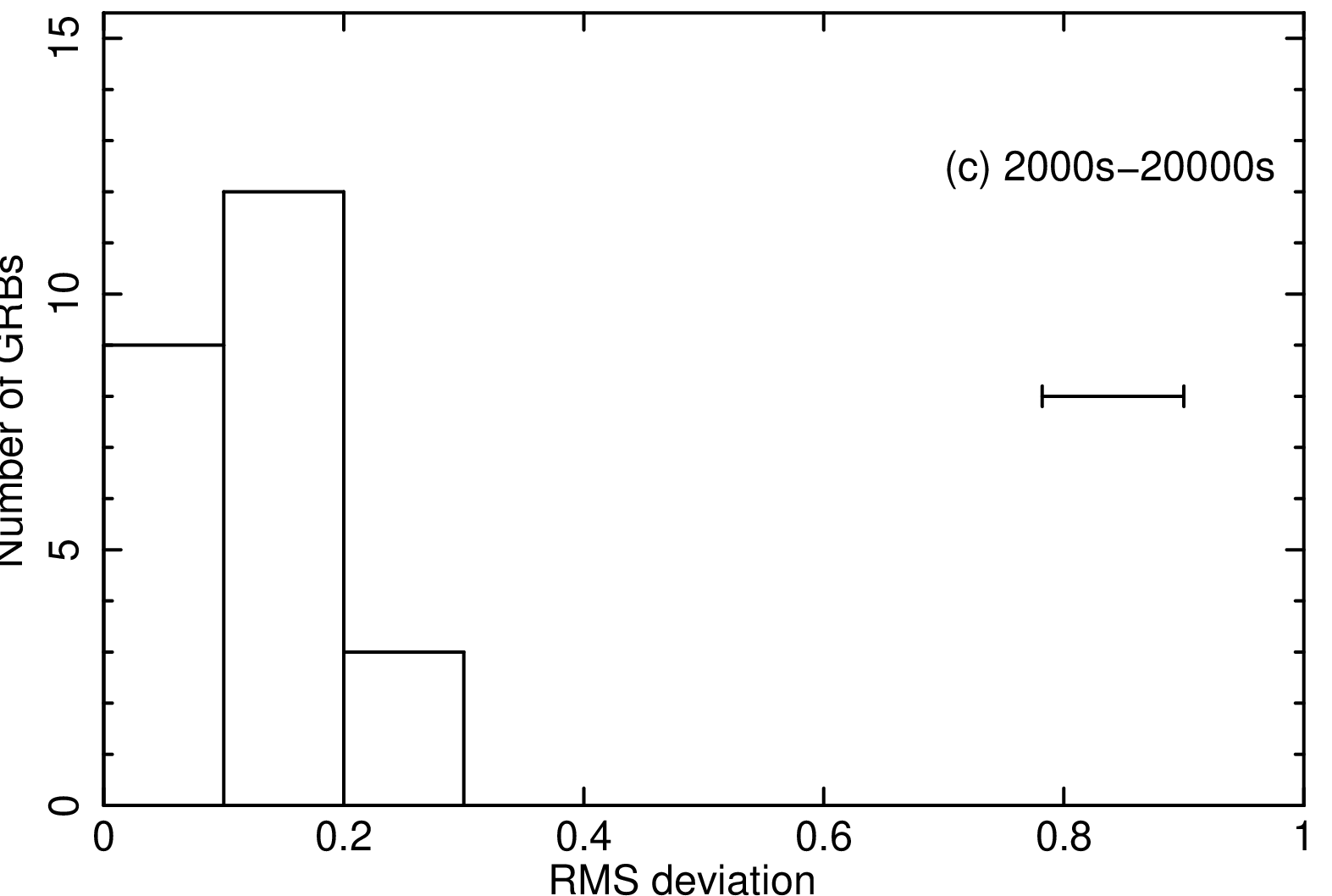}
\vspace{0.2cm}
\includegraphics[angle=-90,scale=0.27]{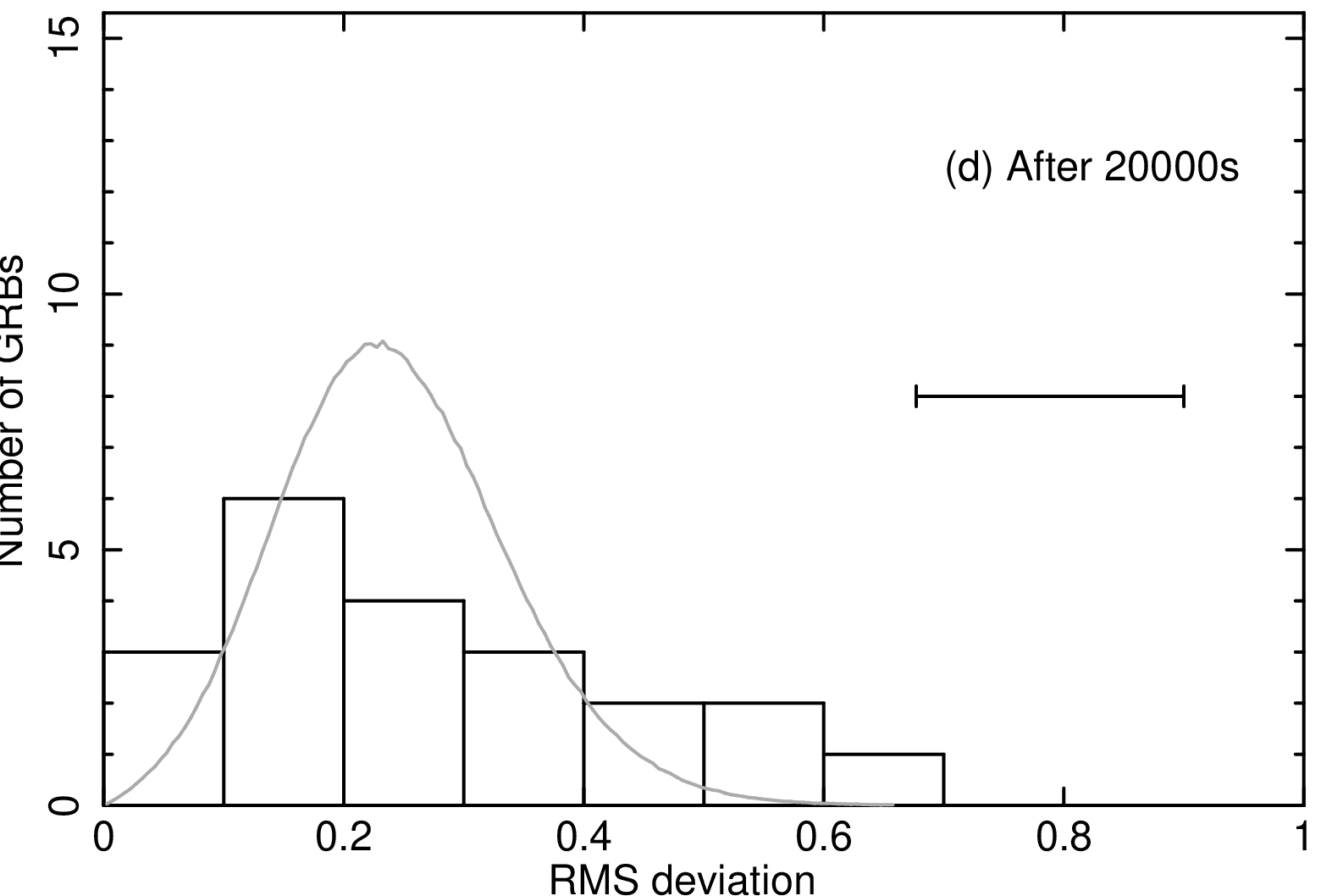}
\caption{Distribution of RMS deviation values determined from the normalized optical/UV and X-ray logarithmic light curves 
during 4 epochs: $<500$s, 500s-2000s, 2000s-20000s and $>20000$s. The error bar in the top right corner represents the 
average error of the RMS deviation in that particular epoch. In panel (d), the histogram shows the distribution of RMS values determined 
from the normalized optical/UV and X-ray logarithmic light curves, while the grey line shows the normalized distribution of the RMS deviation 
values from panel (c) convolved with the mean error from panel (d), see Section \ref{RMS_section} for details.}
\label{correlation}
\end{figure}

The RMS deviation was also determined for each GRB at 4 different specific epochs  (a) to (d), which are $<500$s, 500s-2000s, 2000s-20000s 
and $>20000$s, respectively and are marked on Fig. \ref{light curves}. Histograms of the RMS deviation for epochs (a) to (d) 
can be seen in Fig. \ref{correlation}. The first epoch was selected to end at 500s because by this time the optical/UV afterglows 
have finished rising and the optical/UV light curves have been observed for at least 100s \citep{oates09}. Furthermore, this epoch 
finishes after the first X-ray break in the X-ray light curve, which occurs typically between 200s-400s \citep{eva09}. The second 
epoch was selected to end at 2000s because there is an observing gap between $\sim1000$s and $\sim3000$s. The third epoch starting from 
2000s was chosen to be one dex wide and so ends at 20000s. From 20000s onwards, the signal to noise of the data beings to worsen, particularly 
in the optical, and observations end, with some GRB observations ending as soon as $\sim10^5$s. We therefore took the fourth segment to be from 
20000s until the end of observations because a fifth segment would contain very few GRBs with few optical/UV and X-ray data points.

To allow systematic comparisons of the distribution of RMS deviation with the temporal indices, these 4 epochs were also used when we 
measured the temporal indices of the light curves at multiple epochs. The determination of these values shall be described next. 

\subsection{Temporal Indices}
To determine how the overall behaviour of the optical/UV and X-ray light curves compare over the duration of the observations, 
we fit power-laws individually to the optical/UV and X-ray data that lie within several successive epochs and compared the resulting values. 
The power laws were fitted to the data within the time frames: $<500$s, 500s-2000s, 2000s-20000s and $>20000$s. The best 
fit values were determined using the IDL Levenberg-Marquardt least-squares fit routine supplied by C. Markwardt \citep{mark09}. To ensure 
the power-laws were constrained, the fits were only performed if there were at least 2 data points in both the optical/UV and X-ray 
light curves during the given epoch for which the signal to noise was $>1$. 

Since we are systematically comparing the behaviour of the optical/UV and X-ray light curves, we do not exclude the flaring 
behaviour because it is difficult to do this systematically. Instead we note that the temporal indices may be affected, 
particularly in the early afterglow, due to the presence of flares. Furthermore, as all data in each epoch are fit with 
a power-law, if a break or a flare is present in that epoch the fit will determine a temporal index which corresponds to the 
overall evolution of the light curve, but which does not necessarily correspond to a genuine period of power-law decay. The 
optical/UV and X-ray temporal indices for all four epochs are given in Table \ref{GRBs}. A comparison of the optical/UV 
and X-ray temporal indices for the four time frames are shown in Fig. \ref{decays}. We have also determined the mean 
and intrinsic dispersion of the optical/UV and X-ray temporal indices for each epoch using the maximum likelihood
 method \citep{mac88}, which assumes a Gaussian distribution. These values can be seen in Table \ref{Mean}.

\begin{table*}
\begin {tiny}
\begin{tabular}{|@{}l@{}|c|cccccccc}
\hline
&   & \multicolumn{8}{|c|}{-----------------------------------------Temporal Index---------------------------------------} \\
&   & \multicolumn{4}{|c|}{-----------------------Optical/UV------------------------}& \multicolumn{4}{|c|}{---------------------------X-ray---------------------------} \\
GRB  & Redshift & $<500$s & 500s-2000s & 2000s-20000s & $>20000$s & $<500$s & 500s-2000s & 2000s-20000s & $>20000$s \\
\hline
050319   & $	3.24^{a} $   &0.86  $\pm$1.33 &-0.59$\pm$0.34&-0.48$\pm$0.20&-0.92$\pm$0.16&-7.76$\pm$1.17&-0.70$\pm$0.23  &-0.68$\pm$0.14 &-1.24$\pm$0.11\\
050525   & $	0.606^{b} $  &-1.28 $\pm$0.04 &-0.97$\pm$0.10&-0.91$\pm$0.07&-1.18$\pm$0.09&-0.96$\pm$0.03&-1.13$\pm$0.10  &-1.51$\pm$0.12 &-1.31$\pm$0.09\\
050712	 & 	  -	     &0.10  $\pm$0.64 &-1.25$\pm$0.62&-1.01$\pm$1.37&-0.30$\pm$0.16&-0.64$\pm$0.11&-2.89$\pm$0.33  &-0.63$\pm$0.26 &-1.11$\pm$0.06\\
050726   &	  -	     &-2.67 $\pm$0.80 &-0.71$\pm$3.69&     -        &       -      &-0.17$\pm$0.14&-0.42$\pm$0.67  &      -        &       -      \\
050730   & $	3.97^{c} $   &0.16  $\pm$0.51 &-0.27$\pm$0.88&-0.90$\pm$0.27&-2.17$\pm$0.99&-1.10$\pm$0.08&-1.31$\pm$0.12  &-1.00$\pm$0.06 &-2.67$\pm$0.06\\
050801   & $	1.38^{*}$    &-0.50 $\pm$0.06 &-0.90$\pm$0.21&-0.69$\pm$0.26&       -      &-0.37$\pm$0.21&-1.78$\pm$0.69  &-1.34$\pm$0.20 &        -     \\
050802   & $	1.71^{d} $   &-0.09 $\pm$0.46 &-0.68$\pm$0.10&-0.60$\pm$0.06&-0.81$\pm$0.06&0.75 $\pm$0.28&-0.70$\pm$0.09  &-1.11$\pm$0.04 &-1.42$\pm$0.06\\
050922c  & $	2.198^{e} $  &-1.02 $\pm$0.05 &-0.60$\pm$0.32&-1.04$\pm$0.07&-1.12$\pm$0.12&-0.85$\pm$0.05&-0.92$\pm$0.71  &-1.17$\pm$0.10 &-1.48$\pm$0.17\\
051109a  & $	2.346^{f} $  &-0.52 $\pm$0.44 &     -        &-0.54$\pm$0.12&-0.67$\pm$0.07&-2.80$\pm$0.30&     -          &-1.10$\pm$0.05 &-1.32$\pm$0.03\\
060206   & $	4.04795^{g}$ &-1.89 -    2.18 &     -        &-1.15$\pm$0.17&-1.18$\pm$0.09&1.14$\pm$2.36&     -          &-0.95$\pm$0.10 &-1.36$\pm$0.04\\
060223a  & $	4.41^{h} $   &-0.77 $\pm$0.68 &-0.40$\pm$0.59&     -        &        -     &-0.14$\pm$0.26&4.76$\pm$0.01   &       -       &     -        \\
060418   & $	1.4901^{i} $ &0.01  $\pm$0.03 &-1.39$\pm$0.10&-1.34$\pm$0.09&        -     &-3.21$\pm$0.03&-0.94$\pm$0.19  &-2.29$\pm$0.22 &     -        \\
060512   & $	0.4428^{j} $ &-0.74 $\pm$0.08 &     -        &-0.82$\pm$0.11&-1.53$\pm$0.34&-1.45$\pm$0.11&     -          &-1.21$\pm$0.17 &-1.02$\pm$0.23\\
060526   & $	3.221^{k} $  &-0.31 $\pm$0.08 &-0.20$\pm$0.13&-0.66$\pm$0.75&        -     &1.41 $\pm$0.04&-3.35$\pm$ 0.16 &0.52  -   0.71 &     -        \\
060605   & $	3.8^{l}  $   &0.24  $\pm$0.13 &     -        &-0.86$\pm$0.15&        -     &-1.42$\pm$0.24&     -          &-1.37$\pm$0.09 &     -        \\
060607a  & $	3.082^{m} $  &0.38  $\pm$0.02 &-1.31$\pm$0.06&-1.18$\pm$0.18&        -     &-0.87$\pm$0.02&-0.60$\pm$0.07  &-1.59$\pm$0.07 &     -        \\
060708   & $    1.92^{*}$    &-0.02 $\pm$0.11 &     -        &-0.75$\pm$0.06&-0.98$\pm$0.09&-3.78$\pm$0.10&     -          &-0.80$\pm$0.07 &-1.28$\pm$0.06\\
060804   &        -          &-0.72 $\pm$0.16 &1.70$\pm$2.57 &-0.26$\pm$0.24&-0.33$\pm$0.15 &0.39 $\pm$0.25&-3.77$\pm$1.34 &-1.50$\pm$0.19 &-0.86$\pm$0.21\\
060908   & $	2.43^{n} $   &-1.19 $\pm$0.05 &-1.16$\pm$0.17&-2.18$\pm$0.96&-0.53$\pm$0.37&-0.63$\pm$0.11&-1.07$\pm$0.28  &1.26 -    1.19 &-1.12$\pm$0.19\\
060912   & $	0.937^{o} $  &-0.98 $\pm$0.09 &-1.01$\pm$0.18&-0.59$\pm$0.28&-0.75$\pm$0.18&-0.74$\pm$0.23&-1.10$\pm$0.18  &-1.27$\pm$0.18 &-1.03$\pm$0.19 \\
061007   & $	1.262^{p} $  &-1.69 $\pm$0.11 &-1.70$\pm$0.02&-1.48$\pm$0.03&        -     &-1.83$\pm$0.10&-1.55$\pm$0.02  &-1.75$\pm$0.05 &        -      \\
061021   & $    0.77^{*}$    &-0.93 $\pm$0.06 &     -        &-0.58$\pm$0.05&-1.24$\pm$0.03&-1.83$\pm$0.05&     -          &-0.99$\pm$0.05 &-1.13$\pm$0.01 \\
061121   & $    1.314^{q}$   &-0.12 $\pm$0.05 &-0.80$\pm$0.12&-0.48$\pm$0.09&-0.32$\pm$0.08&-3.90$\pm$0.04&-0.40$\pm$0.05  &-0.99$\pm$0.05 &-1.56$\pm$0.03 \\
070318   & $	0.836^{r} $  &0.42  $\pm$0.03 &-0.96$\pm$0.03&-1.26$\pm$0.08&-0.78$\pm$0.03&-0.23$\pm$0.03&-1.31$\pm$0.11  &-0.92$\pm$0.10 &-1.08$\pm$0.04 \\    
070420	 & $	3.01^{*}  $  &0.72  $\pm$0.14 &-1.94$\pm$0.18&-1.25$\pm$1.35&        -     &-4.38$\pm$0.12&-0.23$\pm$0.10  &-1.24$\pm$0.09 &       -       \\
070529   & $	2.4996^{s} $ &-1.67 $\pm$0.14 & 0.07$\pm$0.57&-0.22$\pm$1.79&-0.62$\pm$0.30&-1.54$\pm$0.23&-1.02$\pm$0.32  &-0.82$\pm$0.60 &-0.96$\pm$0.20 \\
\hline							     
\end{tabular}
\end{tiny}
\caption{Spectroscopic redshifts were largely taken from the literature. For four GRBs, photometric redshifts, indicated by an *, were determined 
using the XRT-UVOT SEDs \citep[see][for details]{oates09}. The table also displays the temporal indices for the optical/UV and X-ray light curves 
for the four epochs: $<500$s, 500s-2000s, 2000s-20000s and $>20000$s. References: a) \protect\cite{jak06} b) \protect\cite{3483} c) \protect\cite{3709} 
d) \protect\cite{3749} e) \protect\cite{jak06} f) \protect\cite{4221} g) \protect\cite{4692} h) \protect\cite{4815} i) \protect\cite{5002} 
j) \protect\cite{5217} k) \protect\cite{jak06} l) \protect\cite{5223} m) \protect\cite{5237} n) \protect\cite{5555} o) \protect\cite{5617} 
p) \protect\cite{5716} q) \protect\cite{5826} r) \protect\cite{6216} s) \protect\cite{6470}.}
\label{GRBs}
\end{table*}

\section{Results}
\label{results}
The XRT and UVOT light curves are shown in Fig. \ref{light curves}. A preliminary examination 
shows that for the majority of GRBs, the optical/UV and X-ray light curves decay at similar rates overall. 
However, there are noticeable differences which tend to be observed at the beginning and 
tail ends of the light curves. For some GRBs (e.g. GRB~060708 and GRB~070318), during the early afterglow, 
the X-ray light curves decay more rapidly than the optical/UV and some of the optical/UV light curves rise. 
This behaviour tends to cease within a few hundred seconds, after which both the optical/UV and X-ray light curves decay at a 
similar rate. For a number of GRBs (e.g GRB~050802 and GRB~060912), towards the end of observations the X-ray light curves appear to 
decay more quickly than the optical/UV light curves. Another noticeable feature is the presence of flares in 
the X-ray afterglows (e.g GRB~060526 and GRB~060607a), which are not often observed in the optical/UV light curves 
and rarely at the same time as those observed in the X-ray light curves. 

In the following 3 subsections, we describe the results of comparing the optical/UV and X-ray light curves 
using the three techniques outlined in Section 2. These three techniques provide information on the similarities 
between the optical/UV and X-ray afterglows in slightly different ways. The hardness ratio provides information 
on how the individual data points behave relative to each other and is a good indicator of temporal changes such 
as breaks in either band, flaring and rising behaviours. The RMS deviation is a good indicator of how well 
the optical/UV and X-ray light curves track each other and the temporal indices determined at the four epochs 
provide information on the average decay rates of the X-ray and optical/UV light curves during the 4 epochs 
(a) to (d) as defined in Section \ref{reduction}. Combining the information from these three techniques enables 
a comprehensive picture to be produced of the X-ray and optical/UV light curves using a systematic and statistical approach.

\subsection{X-ray to Optical/UV Hardness Ratio}
The optical/UV to X-ray hardness ratios are shown in the middle panes of Fig. \ref{light curves}. These hardness 
ratios indicate relative spectral changes between the optical/UV and X-ray light curves, which could be due to 
the passage of a synchrotron frequency through an observed band, differences in the geometries of the emitting 
regions, or due to additional or different emission mechanisms.

For the GRBs in this sample, the hardness ratios exhibit the most rapid variability during the first 1000s, 
after which any changes tend to be more gradual. This corresponds to some of the optical/UV light curves 
rising, some of the X-ray light curves decaying steeply and X-ray flares (e.g GRB~060418 and GRB~060526), 
which all typically occur within the first 1000s. If none of these behaviours are observed within the first 1000s, then the 
hardness ratio is observed to be fairly constant (e.g GRB~050922c and GRB~061007). Periods of constant 
behaviour are an important indication that the X-ray and optical/UV light curves behave the same and that 
the production of emission during this period is intrinsically connected. After the first 1000s, the hardness ratios vary more 
slowly and either are constant, or slowly decrease. For some of the afterglows that have a constant hardness ratio, after a period, 
typically between $\sim2000$s and $\sim10^5$s, the hardness ratio begins to slowly decrease (e.g GRB~050525 and GRB~050802). 
As we do not see the optical/UV light curves change to shallower decays around the same time as the X-ray to optical/UV 
hardness ratios decrease, this implies that the X-ray light curves decay more steeply than the optical/UV light curves 
during this period of hardness ratio decrease. From the hardness ratios alone the reason for the change in X-ray 
temporal index cannot be determined.

\subsection{Root Mean Square Deviation}
\label{RMS_section}
The RMS deviation of the optical/UV and X-ray light curves can be seen for each GRB in the bottom pane of 
each panel of Fig. \ref{light curves}. For most GRBs, there seems to be at least some period where the RMS deviation 
is consistent with zero, indicating similar behaviour in the X-ray and optical/UV light curves for that period. 
For a few GRBs (e.g GRB~050922c and GRB~061007) the RMS deviation is consistent within errors with zero for 
almost their entire duration indicating that the X-ray and optical/UV track each other very well. Roughly 
half the GRBs have RMS deviations, for at least half a dex, that are consistent within errors with having constant RMS deviation, 
but at a value greater than zero, suggesting that the X-ray and optical/UV light curves behave consistently 
different (e.g GRB~060607a and GRB~060804). If the light curves behave consistently different this could 
indicate that the X-ray and optical/UV bands lie either side of a spectral frequency (see Section \ref{discussion}).

A notable period of RMS deviation is before $\sim1000$s, where for a number of GRBs the RMS 
deviation is highly inconsistent with zero and varies rapidly (e.g GRB~060418 and GRB~061121). This early period is where strong differences are 
observed in the behaviour of the optical/UV and X-ray light curves, which is reflected in their RMS deviations. 
Other inconsistencies of the RMS deviation from zero occur at around the same time as apparent changes in the temporal 
index in either the X-ray or optical/UV light curves. For instance GRB~050730 and GRB~050802 both have significant RMS deviations at $\sim3\times10^{4}$s, 
around the time that the X-ray light curve changes decay rate. 

Four histograms were produced using the RMS deviation values determined for each GRB in the time intervals $<500$s, 
500s-2000s, 2000s-20000s and $>20000$s. The number of GRBs in each distribution are 26, 19, 24 and 21, for the 
four epochs respectively. These histograms are shown in Fig. \ref{correlation} and the mean and standard 
deviation of the RMS deviation distributions are given in Table \ref{Mean}. The histogram for $<500$s is shown 
in panel (a). This panel has the widest RMS deviation distribution of all the four panels. The majority of GRBs lie within 0.30, 
but a few produce a tail stretching to 0.90. The distribution narrows by the second panel, which shows the 
RMS deviation values determined from the epoch 500s-2000s, and the GRBs typically have lower RMS deviation values. 
This is also reflected in the lower values for the mean and standard deviation in Table \ref{Mean}. By 2000s-20000s, 
shown in panel (c), the distribution is at its narrowest and the individual RMS deviation values are the lowest of all 
the four epochs, which is also indicated in Table \ref{Mean} by the lowest mean and smallest standard deviation. In panel 
(d), showing the distribution from $>20000$s, the range in RMS deviation values widens. However, the errors on the RMS 
deviation values are also significantly larger at $>20000$s, suggesting that the widening of the distribution could be due 
to the larger uncertainties on the data points at this time. To check this we performed a monte carlo simulation 
of the distribution of the RMS deviation values in panel (c) convolved with the mean error of panel (d). 
To achieve this, for each light curve contributing to panel (d) we perturbed the values of $\log C_{\rm O}-\log C_{\rm X}$ 
in the 2000s-20000s epoch by random displacements drawn from a Gaussian distribution with sigma equal to the 
mean RMS error of panel (d), and computed the resulting RMS. This process was repeated $1 \times 10^{5}$ times for 
each lightcurve to produce the simulated distribution. The normalized, simulated distribution is shown for 
comparison with the real distribution in panel (d). A Kolmogorov Smirnov test comparing the real and simulated 
distributions shown in panel (d) returns a null-hypothesis probability of 28 per cent, implying 
that the distribution in panel (d) could intrinsically be the same distribution as in panel (c), 
but wider due to the larger uncertainty at later times. 

\begin{figure}
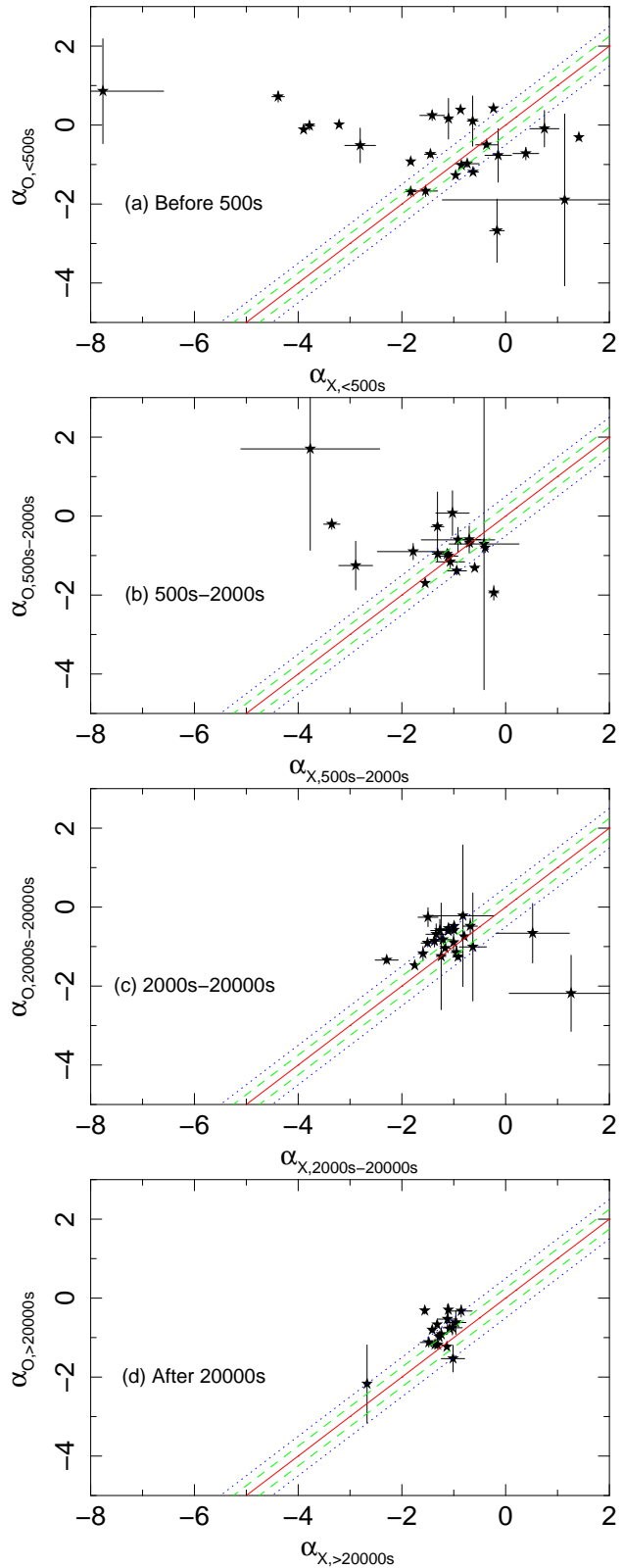

\begin{center}
{\includegraphics[angle=-90,scale=0.33]{Fit_10_500.cps}}
{\includegraphics[angle=-90,scale=0.33]{Fit_500_2000.cps}}
{\includegraphics[angle=-90,scale=0.33]{Fit_2000_20000.cps}}
{\includegraphics[angle=-90,scale=0.33]{Fit_20000_onwards.cps}}
\end{center}
\caption{X-ray and optical/UV temporal indices determined from the light curves during four epochs: $<500$s, 500s-2000s, 2000s-20000s 
and $>20000$s. The red solid line indicates where the optical/UV and X-ray temporal indices are equal. The green dashed lines indicate 
where $\alpha_{O}=\alpha_{X}\pm0.25$ and the blue dotted lines represent $\alpha_{O}=\alpha_{X}\pm0.50$.}
\label{decays}
\end{figure}

\subsection {Comparison of the X-ray and Optical/UV Temporal Indices}
\label{XRT/UVOT_temporal_comparison}
\begin{table*}
\center
\begin{tabular}{|l|cccccc}

\hline
   & \multicolumn{4}{|c|}{------------------Temporal Index------------------ } & \multicolumn{2}{|c|}{------RMS Deviation------} \\
   & \multicolumn{2}{|c|}{-----Optical/UV------} &\multicolumn{2}{|c|}{-----X-ray------} & \multicolumn{2}{|c|}{} \\

Time  & Mean & Dispersion & Mean & Dispersion & Mean & Standard Deviation  \\
\hline
$<500$s      & $-0.51^{+0.17}_{-0.16}$  & $0.67^{+0.19}_{-0.06}$ & $-1.47^{+0.43}_{-0.32}$ & $1.66^{+0.38}_{-0.15}$ & 0.29 & 0.25 \\
500s-2000s   & $-0.98^{+0.14}_{-0.16}$    & $0.42^{+0.16}_{-0.06}$ & $-0.97^{+0.45}_{-0.41}$ & $1.60^{+0.42}_{-0.17}$ & 0.16 & 0.14 \\ 
2000s-20000s & $-0.88^{+0.11}_{-0.08}$    & $0.30^{+0.10}_{-0.04}$ & $-1.15^{+0.07}_{-0.12}$ & $0.32^{+0.11}_{-0.04}$ & 0.12 & 0.05 \\
$>$20000s    & $-0.84\pm0.11$           & $0.31^{+0.11}_{-0.06}$ & $-1.32^{+0.13}_{-0.11}$ & $0.39^{+0.11}_{-0.05}$ & 0.27 & 0.17 \\
\hline							     
\end{tabular}
\caption{For the four epochs, this table provides the mean and intrinsic dispersion  of the temporal indices of the X-ray and 
optical/UV light curves, and the mean and standard deviation of the RMS deviations.}
\label{Mean}
\end{table*}  

\begin{figure}
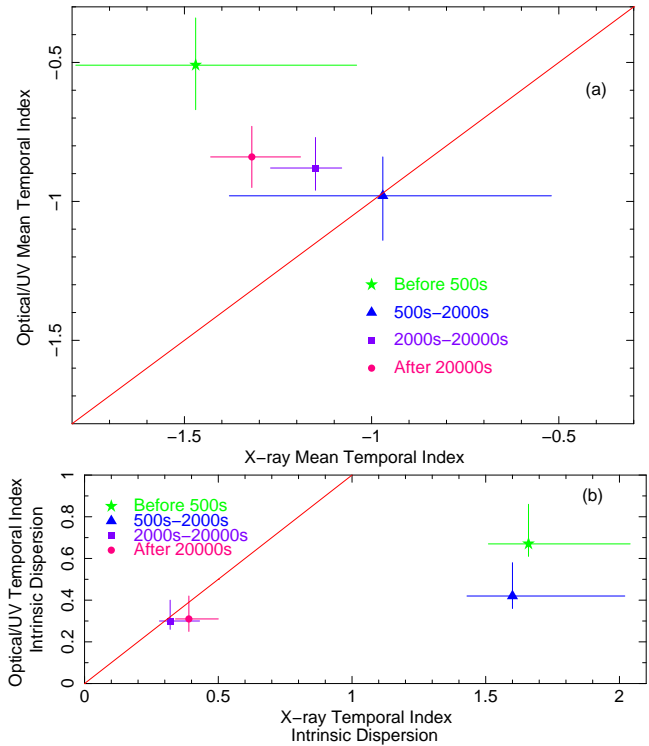

\begin{center}
\includegraphics[angle=-90,scale=0.35]{Mean_fits.cps}
\includegraphics[angle=-90,scale=0.35]{Dispersion_fits.cps}
\end{center}
\caption{Panels (a) and (b) show the mean and intrinsic dispersion, respectively, of the X-ray and 
optical/UV temporal indices at 4 epochs. The red solid line in panel (a), represents the line of equal 
temporal index. The red solid line in panel (b), represents the line of equal intrinsic dispersion.}
\label{Mean_Dispersion} 
\end{figure}

The optical/UV and X-ray temporal indices determined for the epochs: $<500$s, 500s-2000s, 2000s-20000s, 
and $>20000$s are shown in panels (a) to (d) of Fig. \ref{decays}. The individual panels contain 26, 20, 
24 and 17 GRBs, respectively. In each panel of Fig. \ref{decays} the red solid line indicates where the 
optical/UV and X-ray temporal indices, $\alpha_{\rm O}$ and $\alpha_{\rm X}$ respectively, are equal. 
Points lying above the line decay more quickly in the X-ray than in the optical/UV and the points below 
the line decay more quickly in the optical/UV than in the X-ray. The green dashed lines indicate where 
$\alpha_{O}=\alpha_{X}\pm0.25$ and the blue dotted lines represent $\alpha_{O}=\alpha_{X}\pm0.50$, 
where $\Delta\alpha=0.25$ is expected if the synchrotron cooling frequency $\nu_{\rm c}$, lies between 
the X-ray and optical/UV bands and $\Delta\alpha=0.50$ is the maximum difference expected if $\nu_{\rm c}$ 
lies between the X-ray and optical/UV bands and the afterglow is experiencing energy injection. In 
panels (a) and (b) of Fig. \ref{Mean_Dispersion}, we show the X-ray and optical/UV means and intrinsic 
dispersions respectively, for each of the four epochs. An initial examination of Figs. \ref{decays} and 
\ref{Mean_Dispersion} shows that the individual X-ray temporal indices change more than the optical/UV temporal 
indices over the four epochs. This indicates that the change from the scattered distribution of GRBs in the first panel of Fig. \ref{decays} 
to the clustering of the GRBs in the third and fourth panels of Fig. \ref{decays} is predominantly due to the 
change in the X-ray temporal indices.

For the first two epochs, shown in panels (a) and (b) of Fig. \ref{decays}, the GRBs are not tightly clustered 
and appear to have a wide range of X-ray temporal indices, which is also seen in panel (b) of Fig. \ref{Mean_Dispersion} 
as the large intrinsic dispersion in $\alpha_{\rm X}$ for the two epochs before 2000s. In  panels (a) and (b) of Fig. 
\ref{decays}, there are approximately equal numbers of GRBs above and below the line of equal temporal index, 
implying that the optical/UV light curves for some GRBs decay faster than the X-ray light curves, while for 
other GRBs the X-ray light curves decay faster than the optical/UV light curves. A large fraction of GRBs 
in these two epochs have a difference of $\Delta\alpha>0.5$ between the X-ray and optical/UV temporal indices 
implying large differences in the decay of the two bands and indicating that the difference is probably not 
due to the cooling frequency being positioned between the two bands. For the first epoch, shown in panel (a) there are 4 GRBs 
with rising X-ray light curves, indicated by a best fit temporal index of $\alpha_{X,<500s}>0$, and 7 GRBs 
with rising optical/UV light curves indicated by a best fit temporal index of $\alpha_{O,<500s}>0$.

For the last two epochs, given in panels (c) and (d) of Fig. \ref{decays}, the majority of the light curves 
are quite tightly clustered, implying that most of the GRB afterglows behave similarly at late times. The 
narrow range in temporal indices can also be observed in Fig. \ref{Mean_Dispersion}, by the small values of 
intrinsic dispersion of both the optical/UV and X-ray temporal indices. In both epochs, only a small number 
of GRBs have differences between optical/UV and X-ray temporal indices of $\Delta\alpha>0.5$. More importantly, the 
majority of the GRBs in the last two epochs lie above the line of equal temporal index, implying that the 
optical/UV light curves decay more slowly than the X-ray light curves. One possible cause of a shallow decay 
in the optical/UV light curves would be a strong contribution from the host galaxy. If the host galaxy 
contribution was significant then at the tail end of the optical/UV light curve a constant count rate would 
be observed. However, for the majority of GRBs in this sample we do not observe a flattening at late times, 
implying that the optical/UV contribution from the host galaxy has a negligible effect on the light curve, and 
is not the reason why the optical/UV light curves decay on average less steeply compared with the X-ray light 
curves. The trend that the optical/UV light curves decay more slowly than the X-ray light curves is also 
indicated in Fig. \ref{Mean_Dispersion}, with the mean temporal indices for the epochs 2000s-20000s and 
$>20000$s sitting above the line of equal temporal index. In fact even for the first two epochs, the mean 
values lie above or are consistent with lying above the line of equal temporal index, suggesting that X-ray 
light curves decay faster on average than optical/UV light curves throughout the entire observing period. 
Furthermore, for the epoch $>20000$s, shown in panel (d) of Fig. \ref{decays}, the GRBs are clustered slightly 
to the left of those of the previous epoch 2000s-20000s, shown in panel (c). This can also be seen in Fig. 
\ref{Mean_Dispersion}, with the X-ray mean for the $>20000$s at a slightly lower value than the 2000s-20000s mean. 
This suggests that at least for some GRBs, there is a change in the X-ray temporal index to steeper values. This 
was also suggested in Section 3.1 from investigating the hardness ratios.

It is not possible, when investigating panels (c) and (d) of Fig. \ref{decays} individually, to determine how many 
light curves display a change in X-ray or optical/UV temporal index. Therefore, we have determined in Table \ref{alpha_difference} 
for the 17 GRBs in panel (d), the difference between the X-ray and optical/UV temporal indices determined at both 
the 2000s-20000s and $>20000$s epochs. The table is coded by three symbols which divides the GRBs by temporal behaviour: 
both the X-ray and optical/UV temporal indices become more negative (triangles); the X-ray temporal index become more 
negative, but the optical/UV temporal index becomes more positive (squares); and the X-ray temporal index becomes more positive, but the 
optical/UV temporal index becomes more negative (circles). The first thing to note is that there are no GRBs whose 
X-ray and optical/UV light curves both become shallower in the $>20000$s epoch. 

The most common behaviour, which occurs for 9 of the 17 GRBs, is that both the best fit X-ray and optical/UV temporal 
indices become more negative i.e both light curves become steeper. For the rest of the GRBs, 4 become steeper in the 
optical/UV, but shallower in the X-ray and 4 become steeper in the X-ray, while becoming shallower in the optical/UV. 
Examining the significance of the changes to these 17 GRBs we find that 7 GRBs, GRB~050319, GRB~050730, GRB~051109a, 
GRB~060206, GRB~060804 GRB~060908 and GRB~061121 are consistent with no change in the optical/UV temporal index, while 
the X-ray is inconsistent at $\geq2\sigma$, indicating a break. The hardness ratios of these GRBs provides evidence 
that the breaks are chromatic because the hardness ratios soften for these GRBs during the last two epochs 
(see also  Section 3.1). GRBs, GRB~050525, GRB~060512, and GRB~070318 have X-ray temporal indices that are consistent 
with no change, while the optical/UV temporal index between the two epochs is not consistent with being the same at 
$\geq2\sigma$, which suggests a chromatic break. However, the hardness ratio of GRB~050525 does not show an obvious 
hardening, which would be expected if a break was observed in the optical and not the X-ray, but it does soften in 
the 2000s-20000s epoch and becomes constant during the $>20000$s epoch. The hardness ratios for GRB~060512 
and GRB~070318 appear to be constant during the last two epochs, implying that there is not a break in the optical/UV. 
GRBs, GRB~050712, GRB~050922c, GRB~060912 and GRB~070529 are consistent with no change in either the X-ray or the 
optical/UV and the remaining 3 GRBs have optical/UV and X-ray temporal indices that are different between the two 
epochs at $\geq2\sigma$, suggesting a change in temporal index in both light curves.

\begin{figure}
\begin{center}
\includegraphics[angle=-90,scale=0.38]{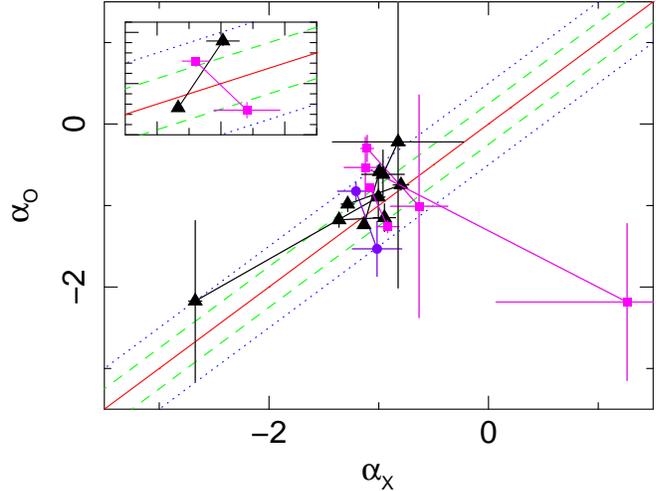}
\end{center}
\caption{This figure plots the X-ray and optical/UV temporal indices determined from the epochs 2000s-20000s and 
$>20000$s for 8 GRBs, which appear to cross the line of equal temporal index. Each GRB has a pair of data points 
linked together which show the temporal indices in the 2000s-20000s and $>20000$s epochs. The symbols, which 
correspond to those in Table \ref{alpha_difference}, show how the temporal index changes between the two epochs. 
Pairs of black triangles are those GRBs for which the X-ray and optical/UV temporal indices become more negative 
between the epochs 2000s-20000s and $>20000$s (i.e they move down and to the right). Pairs of purple circles are those 
GRBs in which the X-ray temporal index becomes less negative and the optical/UV temporal index becomes more negative 
(i.e they move down and to the left) and the pairs of pink squares are GRBs for which the X-ray temporal index becomes 
more negative and the optical/UV temporal index becomes less negative (i.e they move up and to the right). The red 
solid line indicates where the optical/UV and X-ray temporal indices are equal, the green dashed line indicates 
where $\alpha_{O}=\alpha_{X}\pm0.25$ and the blue dotted lines represents $\alpha_{O}=\alpha_{X}\pm0.50$. The inserted 
panel shows the two GRBs, GRB~070318 and GRB~061021, which are consistent with having one data point more than 
2$\sigma$ above the line of equal temporal index and the other data point more than 2$\sigma$ below the line of equal temporal index.}

\label{delta}
\end{figure}

The other interesting behaviour, shown in Fig. \ref{delta}, is that a small number of GRBs appear to cross the line 
of equal temporal index, but this is only significant for two GRBs, GRB~070318, GRB~061021. These GRBs have one data 
point more than 2$\sigma$ above the line and the other data point more than 2$\sigma$ below the line of equal temporal index, 
which can be seen in the inset panel of Fig. \ref{delta}. For all other GRBs, at least one of their data points are 
consistent within $2\sigma$ with lying on either side of the line of equal temporal index index. GRB~061021 is consistent 
with crossing from above to below the line of equal temporal index, indicating that a change from the optical/UV 
light curve having a shallower decay than the X-ray to the X-ray light curve having a shallower decay than the optical/UV. 
This is also observed in the hardness ratio for this GRB, which softens during the 2000s-20000s epoch, indicating 
that the X-ray decays more steeply than the optical/UV, but hardens during the $>20000$s epoch indicating that the 
optical/UV light curve decays more rapidly than the X-ray. As for the other GRB, GRB~070318, this GRB is consistent 
with crossing from below to above the line of equal temporal index, indicating a change from the X-ray light curve 
having a shallower decay than the optical/UV to the optical/UV light curve having a shallower decay than the X-ray. 
A subtle softening of the hardness ratio for GRB~070318, implies that the X-ray lightcurve in the $>20000$s epoch 
decays more quickly than the optical/UV.

From Table \ref{alpha_difference} and Fig. \ref{delta} we can draw three significant conclusions: 7 of the 17 ($\sim41\%$) 
afterglows have a break, which is observed only in the X-ray light curve between 2000s and the end of observations; there are no 
afterglows that become shallower in the optical/UV and in the X-ray; 2 GRBs traverse the line of equal temporal index, 
one from above to below the line of equal temporal index and the other from below to above the line of equal temporal index.

\begin{table}
\begin{center}
\begin{small}
\begin{tabular}{l|c|c|c}
\hline
GRB  &  &$\Delta\alpha_X$ & $\Delta\alpha_O$ \\
\hline
GRB050319 & $\blacktriangle$ &  $-0.56\pm0.18$  & $-0.44\pm0.26$\\
GRB050525 & $\bullet$        &  $ 0.20\pm0.15$  & $-0.27\pm0.12$\\
GRB050712 & $\blacksquare$   &  $-0.47\pm0.27$  & $ 0.71\pm1.38$\\
GRB050730 & $\blacktriangle$ &  $-1.67\pm0.09$  & $-1.28\pm1.02$\\
GRB050802 & $\blacktriangle$ &  $-0.31\pm0.07$  & $-0.20\pm0.09$\\
GRB050922c& $\blacktriangle$ &  $-0.32\pm0.19$  & $-0.08\pm0.14$\\
GRB051109a& $\blacktriangle$ &  $-0.22\pm0.06$  & $-0.13\pm0.14$\\
GRB060206 & $\blacktriangle$ &  $-0.41\pm0.11$  & $-0.03\pm0.19$\\
GRB060512 & $\bullet$        &  $0.19 \pm0.28$  & $-0.71\pm0.36$\\
GRB060708 & $\blacktriangle$ &  $-0.49\pm0.09$  & $-0.23\pm0.11$\\
GRB060804 & $\bullet$        &  $0.64 \pm0.29$  & $-0.07\pm0.29$\\
GRB060908 & $\blacksquare$   &  $-2.38\pm1.21$  & $ 1.65\pm1.03$\\
GRB060912 & $\bullet$        &  $0.24 \pm0.26$  & $-0.16\pm0.34$\\
GRB061021 & $\blacktriangle$ &  $-0.14\pm0.05$  & $-0.65\pm0.05$\\
GRB061121 & $\blacksquare$   &  $-0.56\pm0.06$  & $ 0.17\pm0.12$\\
GRB070318 & $\blacksquare$   &  $-0.16\pm0.11$  & $ 0.48\pm0.08$\\
GRB070529 & $\blacktriangle$ &  $-0.14\pm0.63$  & $-0.40\pm1.82$\\
\hline							     
\end{tabular}
\end{small}
\caption{Differences in temporal index from the 2000s-20000s and 20000s onwards epochs. 
Symbols correspond to those in Figure \ref{delta}: $\blacktriangle$ both the X-ray and 
optical become steeper, $\blacksquare$ X-ray becomes steeper while the optical becomes 
shallower, $\bullet$ X-ray becomes shallower while the optical becomes steeper.}
\label{alpha_difference}
\end{center}
\end{table}

\section{Discussion}
\label{discussion}

All three analysis methods indicate that the X-ray and optical/UV light curves behave most differently before 500s. The 
RMS deviation distribution and the mean temporal indices together indicate that the optical/UV and X-ray light curves behave 
most similarly during the 2000s-20000s epoch. For all four epochs we find that the optical/UV light curves decay more 
slowly on average than the X-ray. We also find through investigation of the temporal indices and the hardness ratios 
that chromatic breaks are observed in some of the GRB afterglows, with the breaks observed in the X-ray light curves.

In the following sections we shall examine two models, a single component jet and a jet with additional emission 
regions such as a two component jet, or late `prompt' emission, to determine whether either of these models 
can explain the observations.

\subsection{Single Component Outflow} 

The temporal indices expected from a synchrotron dominated outflow are determined by a set of equations 
\citep[][see last reference for a comprehensive list]{sari98,mes98,sar99,che00,dai01,racusin09}. These are 
mathematical expressions that relate the temporal index to predominantly a micro-physical parameter, a 
physical parameter and the positioning of two spectral frequencies relative to the observed band. Specifically these are 
the electron energy index $p$, which is typically between 2 and 3 \citep{pan02,sta08,cur09}, the density profile 
of the external medium (constant or wind-like) and the relative positions in the spectrum of the synchrotron 
frequencies, primarily the synchrotron cooling frequency $\nu_{\rm c}$ and the synchrotron peak frequency 
$\nu_{\rm m}$ (see Table \ref{Closure_relations}). There is a third synchrotron frequency, the synchrotron 
self-absorption frequency, but this frequency does not influence the optical/UV or the X-rays during the 
timescales studied here. Recent observations of {\it Swift} GRB afterglows have shown that, in some cases, 
the temporal indices are shallower than expected \citep{nousek}. This led to the hypothesis that, at 
least for a certain time, the ejecta may be injected with some additional energy \citep[see][for a discussion 
and for other possible interpretations]{zhang06}. The temporal indices of these GRBs should then be satisfied 
by the energy injected temporal relations (see Table \ref{Closure_relations}). The amount of energy injection 
is measured by the luminosity index, $q$, which varies between 0 and 1 in the luminosity relation $L(t)=L_0(t/t_0)^{-q}$ 
where $t$ is the observers time, $t_0$ is the characteristic timescale for the formation of a self-similar solution, 
which is roughly equal to the time at which the external shock starts to decelerate \citep{zhang01}. When $q=1$ the 
injected temporal relations reduce to the non-injected closure relations. 

The position of the synchrotron cooling frequency relative to the synchrotron peak frequency dictates whether 
electrons are in a slow cooling ($\nu_m<\nu_c$) or fast cooling regime ($\nu_c<\nu_m$). The fast cooling 
closure relations provided in Table \ref{Closure_relations} are valid only in the adiabatic regime and are 
not valid for radiative evolution \citep{sari98}. For a single component jet it is expected that the optical/UV 
and X-ray emission are produced within the same region and therefore are explained by the same synchrotron 
spectrum, with the possibility that one or more of the synchrotron frequencies are between these two observing 
bands. This means that the optical/UV and X-ray temporal indices, determined from an afterglow, should be 
described by temporal relations which rely on the same assumptions about the external medium, the electron 
energy index, $p$, and the value of $q$. 

In order to assess the validity of this scenario, we shall consider the mean X-ray and optical/UV temporal indices 
and search for a common set of closure relations that are allowed in all the four temporal epochs and we shall 
use the hardness ratios and the RMS deviations to support and confirm our findings. To determine if a scenario 
is acceptable, consistent values of $p$ must be derived from the mean temporal indices of the optical/UV and 
X-ray light curves at the different epochs. As we are using the average properties of the sample, we note 
that the conclusion drawn is typical of the sample, but should not be taken as the conclusive explanation 
for individual GRBs, which should be investigated individually. 

\begin{table*}
\begin{center}
\begin{tabular}{|c|c|c||c|ccc}
\hline
 & \multicolumn{2}{c}{Temporal Relations}          &  \multicolumn{2}{c}{$p=2$}  &  \multicolumn{2}{c}{$p=3$}            \\           
          &  Non-injected       &   Energy Injected       &  $q=0$ & $q=1$  &  $q=0$ & $q=1$                        \\    
\hline
\hline
          &        ($q=1$)                       &     ($0\leq q<1$)      &  $\alpha$ & $\alpha$ & $\alpha$ & $\alpha$ \\ 
\hline                                                                                          
\multicolumn{2}{c}{ISM slow cooling}     &                            &          &           &          &           \\
$\nu<\nu_{\rm m}$           &   1/2       & $(8-5q)/6$                 &  1.33    &  0.50     &  1.33    &  0.50     \\       
$\nu_{\rm m}<\nu<\nu_{\rm c}$& $3(1-p)/4$  & $\frac{(6-2p)-(p+3)q}{4}$  &   0.5    & -0.75     &  0.00    & -1.50     \\  
$\nu>\nu_{\rm c}$	   & $(2-3p)/4$  & $\frac{(4-2p)-(p+2)q}{4}$  &  0.00    & -1.00     & -0.50    & -1.75     \\   
\hline                                                                                                  
\multicolumn{2}{c}{ISM fast cooling}     &                            &          &           &          &           \\ 
$\nu<\nu_{\rm c}$	   &    1/6      & $(8-7q)/6$                 &  1.33    &  0.17     &  1.33    &  0.17     \\  
$\nu_{\rm c}<\nu<\nu_{\rm m}$&  -1/4       & $(2-3q)/4$                 &  0.50    &  -0.25    &  0.50    &  -0.25    \\ 
$\nu>\nu_{\rm m}$           & $(2-3p)/4$  & $\frac{(4-2p)-(p+2)q}{4}$  &  0.00    &  -1.00    & -0.50    & -1.75     \\ 
\hline                                                                                                  
\multicolumn{2}{c}{Wind slow cooling}    &                            &          &           &          &           \\  
$\nu_<\nu_{\rm m}$          &    0        & $(1-q)/3$                  &  0.33    &  0.00     &  0.33    &  0.00     \\ 
$v_{\rm m}<v<v_{\rm c}$      & $(1-3p)/4$  & $\frac{(2-2p)-(p+1)q}{4}$  &  0.00    &  -1.25    & -1.00    & -2.00     \\  
$\nu>\nu_{\rm c}$	   & $(2-3p)/4$  & $\frac{(4-2p)-(p+2)q}{4}$  &  0.00    &  -1.00    & -0.50    & -1.75     \\ 
\hline		                                                                                 
\multicolumn{2}{c}{Wind fast cooling}    &                            &          &           &          &           \\  
$\nu<\nu_{\rm c}$	   &    -2/3     & $-(1+q)/3$                 & -0.33    &  -0.67    & -0.33    & -0.67     \\ 
$\nu_{\rm c}<\nu<\nu_{\rm m}$&    -1/4     & $(2-3q)/4$                 &  0.50    &  -0.25    &  0.50    &  -0.25    \\ 
$\nu>\nu_{\rm m}$	   & $(2-3p)/4$  & $\frac{(4-2p)-(p+2)q}{4}$  &  0.00    &  -1.00    & -0.50    &  -1.75    \\ 
\hline		
\multicolumn{2}{c}{Jet slow cooling}     &                            &          &           &          &           \\  
$\nu<\nu_{\rm m}$           &    -1/3     &     -                      &    -     &   -0.33   &    -     & -0.33     \\
$\nu_{\rm m}<\nu<\nu_{\rm c}$& $\sim-p$    &     -                      &    -     &   -2.00   &    -     & -3.00     \\
$\nu>\nu_{\rm c}$	   & $\sim-p$    &     -                      &    -     &   -2.00   &    -     & -3.00    \\
\hline
\end{tabular}
\end{center}
\caption{This table provides the ranges in temporal index for the temporal relations that 
are expected from synchrotron emission with and without energy injection \protect\citep{zhang04,zhang06}. The electron energy index $p$ 
dictates the range of values of the temporal index for each temporal relation. The electron energy index $p$ and the luminosity index $q$ 
dictate the range of values of the temporal index for each temporal relation. When $q=1$ the energy injected temporal relations 
reduce to those of the non-injected cases. The temporal relations for the jet case can be found in \protect\cite{pan06b}.}
\label{Closure_relations} 
\end{table*}

\subsubsection{Implications of the Mean GRB temporal properties}
Before 500s, the mean temporal indices of the X-ray and optical/UV afterglows are $\alpha_{X,<500s}=-1.47^{+0.43}_{-0.32}$ 
and $\alpha_{O,<500s}=-0.51^{+0.17}_{-0.16}$. The X-ray mean temporal index can be explained by several of the non-injected temporal 
relations in Table \ref{Closure_relations}, and the optical/UV mean temporal index can be explained either 
in a scenario with  $\nu<\nu_{\rm c}$, a wind medium and fast cooling electrons (which would be contrived as 
the theoretical temporal index for this scenario is a single distinct value while in reality the optical/UV 
light curves, before 500s have a range in temporal index), or by several of the energy injected temporal relations. 
The lack of discrimination of the temporal expressions for the light curves before 500s is not unexpected as there 
is a wide range in temporal behaviour in both the optical/UV and the X-ray. The wide temporal behaviour is also 
observed in the RMS deviation histogram as a wide distribution during this epoch.

Moving to the next epoch between 500s-2000s, the mean temporal indices of the X-ray and optical/UV light curves are 
$\alpha_{O,500s-2000s}=-0.98^{+0.16}_{-0.14}$ and $\alpha_{X,500s-2000s}=-0.97^{+0.45}_{-0.41}$. Both are consistent with 
the non-injected temporal relations for a slow cooling ISM-like medium with $\nu_{\rm m}<\nu<\nu_{\rm c}$. This gives two values of $p$, 
$p=2.30^{+0.16}_{-0.14}$ from the optical/UV and $p=2.29^{+0.45}_{-0.41}$ from the X-ray, which are consistent to 
within 1$\sigma$.  However, both the X-ray and optical/UV mean temporal indices could also be reproduced by the 
relations for both the ISM-like and wind-like media in the slow cooling case $\nu>\nu_{\rm c}$ and in the fast 
cooling case $\nu>\nu_{\rm m}$ giving values $p=1.97^{+0.16}_{-0.14}$ for the optical/UV and $1.96^{+0.45}_{-0.41}$ for 
the X-ray, which again are consistent to within 1$\sigma$. Furthermore, there is one more option: with slow cooling 
electrons in an ISM-like medium, the values of the temporal indices allow the possibility that $\nu_{\rm m}<\nu_{\rm O}<\nu_{\rm c}<\nu_{\rm X}$, 
which produces values of $p=2.29^{+0.45}_{-0.41}$ for the X-ray and $p=1.97^{+0.16}_{-0.14}$ from the optical/UV, which 
are consistent to within 1$\sigma$. The temporal indices can also be explained by the energy injected relations, in 
these cases the values of $p$ may change depending upon the energy injection parameter $q$. If energy injection is 
considered then the temporal relations for a wind-like medium are also acceptable for slow cooling with either 
$\nu_{\rm m}<\nu_{\rm O}<\nu_{\rm X}<\nu_{\rm c}$ or $\nu_{\rm m}<\nu_{\rm O}<\nu_{\rm c}<\nu_{\rm X}$ and for fast cooling 
with $<\nu_{\rm c}<\nu_{\rm m}<\nu_{\rm O}<\nu_{\rm X}$. The narrower RMS distribution histogram compared to the previous 
epoch, indicates that the optical/UV and X-ray light curves for a large fraction of GRBs behave in a similar way, 
consistent with the expectations of a single synchrotron spectrum producing both light curves.  

During the epoch 2000s-20000s, the mean X-ray temporal index is $\alpha_{X,2000s-20000s}=-1.15^{+0.07}_{-0.12}$ and the mean optical/UV 
temporal index is $\alpha_{O,2000s-20000s}=-0.88^{+0.11}_{-0.08}$. The difference in $\alpha$ between the optical/UV and X-ray indices, 
$\Delta\alpha=0.27^{+0.16}_{-0.10}$, implies that the optical/UV and X-ray do not lie on the same spectral segment. This difference 
is consistent with a cooling break ($\Delta\alpha=0.25$) lying in between the X-ray and optical/UV bands. The 
only non-energy injected temporal relations that can produce both mean values are the ISM slow cooling temporal relations 
for the case $\nu_{\rm m}<\nu_{\rm O}<\nu_{\rm c}<\nu_{\rm X}$. These relations give consistent values of $p$: $p=2.17^{+0.11}_{-0.08}$ 
determined using the optical/UV temporal mean and $p=2.20^{+0.07}_{-0.12}$ determined with the X-ray temporal mean. Looking at the 
temporal values in Table \ref{Closure_relations}, the only temporal relation for a wind-like medium that could explain wide ranges in 
both temporal indices and with the X-ray and optical/UV having different temporal indices would be for the slow cooling case with 
$\nu_{\rm m}<\nu_{\rm O}<\nu_{\rm c}<\nu_{\rm X}$. However, this cannot explain these temporal indices even with energy injection, 
since in the wind-medium (if $\nu_c<\nu_O,\nu_X$ in the fast cooling case or $\nu_m<\nu_O,\nu_X$ in the slow cooling case) the X-ray 
is required to be shallower than the optical/UV by 0.25, which is the opposite of what is observed. As there are only a small number 
of GRBs with a break in the optical/UV light curve \citep{oates09}, the temporal indices are consistent with an ISM-like medium with 
$\nu_{\rm c}$ being between the X-ray and optical/UV bands during the 500s-2000s and 2000s-20000s epochs. This implies that we have 
slow cooling electrons in an ISM-like medium with $\nu_{\rm m}<\nu_{\rm O}<\nu_{\rm c}<\nu_{\rm X}$ from 500s to 20000s. The narrowness 
and the low values of the RMS deviation histogram for the 2000s-20000s epoch, agrees with a single synchrotron spectrum producing 
the X-ray and optical/UV emission for almost all GRBs during this epoch. 

For the final epoch $>20000$s, the mean X-ray temporal index is $\alpha_{X,>20000s}=-1.33^{+0.13}_{-0.11}$ and the 
mean optical/UV temporal index is $\alpha_{O, >20000s}=-0.84\pm0.11$. Again the mean values can only be produced 
by the non-injected temporal relations for the ISM slow cooling regime with $\nu_{\rm m}<\nu_{\rm O}<\nu_{\rm c}<\nu_{\rm X}$. 
Wind-like density cannot explain the temporal indices of this epoch either since, similar to the previous epoch, the 
optical/UV and X-ray temporal indices have wide ranges, but the optical/UV is shallower than the X-ray, which cannot be 
explained by the temporal relations for a wind-like medium, even including energy injection. The values of $p$ determined from the 
non-injected temporal relations for the ISM slow cooling regime with $\nu_{\rm m}<\nu_{\rm O}<\nu_{\rm c}<\nu_{\rm X}$ are 
$p=2.12\pm0.11$ for the optical/UV and $p=2.44^{+0.13}_{-0.11}$ for the X-ray. These values are marginally 
consistent with each other at $2\sigma$. The $p$ value determined from the optical/UV is consistent with 
$p$ value determined from the optical/UV in the previous epoch. The $p$ value determined from the X-ray 
is marginally consistent at 2$\sigma$ with the $p$ value derived from the mean X-ray temporal index from 
the same regime in the previous epochs. The large errors on the RMS deviations determined for the $>20000$s epoch means 
that little can be implied from this $>20000$s RMS deviation distribution. 

The narrowness and the small valued RMS deviation distribution in the 500s-2000s and 2000s-20000s epochs support the hypothesis 
of a single synchrotron emission spectrum from a single component emission region. The general consistency of the mean temporal 
indices with the non-injected temporal relations, producing consistent and realistic $p$ values, suggests that at least from 500s, 
the sample on average is consistent with slow cooling electrons in a constant density medium with $\nu_{\rm m}<\nu_{\rm O}<\nu_{\rm c}<\nu_{\rm X}$. 
The mean temporal indices of the last 3 epochs are consistent with a single component outflow, without the need for energy injection, 
although we cannot exclude the requirement of energy injection, which would complicate this simplistic picture and would increase 
the value of $p$. However, it is unlikely that this simple picture can explain all GRBs and we need to determine how this picture 
changes on a GRB to GRB basis. Therefore, we shall compare this picture with the individual temporal indices at each epoch.

\subsubsection{Implications of the Individual GRB Properties}
In Figure \ref{decays}, in each panel the green dashed line represents the difference between the optical/UV and
 X-ray temporal indices $\Delta\alpha=0.25$, expected when $\nu_{\rm c}$ lies between these bands, and the blue dotted line represents the 
maximum difference $\Delta\alpha=0.5$, expected when $\nu_{\rm c}$ lies between these bands and the afterglow is energy-injected. 
Furthermore, $\alpha_{X}+0.25\leq\alpha_{\rm O}\leq\alpha_{X}+0.50$ is expected for a constant density medium, while 
$\alpha_{X}-0.50\leq\alpha_{\rm O}\leq\alpha_{X}-0.25$ is expected for a wind-like medium.

For the epoch $<500$s, shown in panel (a) of Fig. \ref{decays}, it is clear that the mean temporal indices 
are not representative of the full behaviour of the optical/UV and X-ray light curves. It is also clear from the rapid 
variability in the hardness ratios of individual GRBs and from the changes in RMS deviations during this epoch that a simple 
outflow ploughing into a constant density medium is too simplistic. This is also shown in Fig. \ref{decays} by 
the lack of consistency with $\alpha_{X}\leq\alpha_{O}\leq\alpha_{X}+0.50$. Instead, this figure shows a wide range in behaviour that 
physically can be divided into several groups. 
\begin{itemize}

\item Five GRBs are consistent with $\alpha_{O}=\alpha_{X}-0.25$, suggesting that these GRBs lie in a wind 
medium with a cooling break between the X-ray and optical/UV bands. The temporal range of these GRBs is 
$-1.54<\alpha_{X,<500s}<-0.14$ and  $-1.67<\alpha_{O,<500s}<-0.50$. As the shallowest temporal index 
produced by a wind medium with $\nu_{\rm O}<\nu_{\rm c}<\nu_{\rm X}$ is $\alpha_{\rm X}=\alpha_{\rm O}+0.25=-1.00$, 
this implies that for at least a couple of these GRBs energy injection is required. 

\item Four GRBs have $0<\alpha_{X,<500s}$. A visual inspection of these GRBs during this period, reveals that three of these GRBs have 
flares in the X-ray emission that are not observed in the optical/UV, implying late time central engine activity \citep{fal07}. 
Furthermore, the three GRBs with X-ray flares all have optical/UV temporal indices $-0.80<\alpha_{O,<500s}\lesssim 0.00$, which are too 
shallow to be explained by the non-energy injected temporal relations, therefore, implying energy injection. This 
scenario was also found to be the case for the short-hard GRB~060313 \citep{rom06}. 

\item Six GRBs sit within $-9<\alpha_{\rm X}<-2$, with five sat between $-4.50<\alpha_{\rm X}<-2$. The sixth GRB, GRB~050319 has 
large errors on both the X-ray and optical/UV temporal indices as only two data points fall in the $<500$s epoch. Steep 
decays, such as observed for the 5 other GRBs ( $-4.50<\alpha_{\rm X}<-2$), are expected from the tail of the prompt 
emission \citep{zhang06}, suggesting that the X-ray emission of these 5 GRBs is dominated by prompt emission. These 
GRBs also have RMS deviations that are inconsistent with being zero and hardness ratios that vary rapidly during this 
epoch, which suggests another jet component or another emission component and so lends support to prompt emission 
contaminating the X-ray emission. These GRBs are the only GRBs in the sample with X-ray light curves that appear 
to decay with three of the four segments of the canonical X-ray light curves: an initial steep decay followed by 
the shallow decay and followed finally by a normal decay. For these GRBs, it appears that as the X-ray temporal 
index tends to more negative values, the optical/UV temporal index tends to more positive values. However, with 
only 5 GRBs, we can not determine if the X-ray and optical/UV temporal indices of these GRBs are statistically 
correlated. A larger sample will be required to investigate if a correlation exists.

\item Five GRBs lie between $-2<\alpha_{X,<500s}<0$, but have $\alpha_{O,<500s}>0$. These GRBs are 
rising in the optical/UV during this early epoch. This behaviour can also be observed by the 
varying hardness ratios and the inconsistency of the RMS deviations with zero for three of these GRBs. 
For the other two GRBs the rising behaviour is not observed as clearly as the other GRBs and this is reflected 
in their hardness ratios and RMS deviations. In \cite{oates09}, the rising behaviour was best explained as to be 
due to the start of the forward shock. This should be an achromatic effect and therefore should also be observed 
in the X-ray light curves. Instead what we see is $-2<\alpha_{X,<500s}<0$, which is usually expected for a 
light curve after the start of the forward shock. However, from this analysis it is not possible to determine if the 
rise is masked due to a contribution from the tail of the prompt emission \citep{zhang06} or whether more complex 
jet geometry is required for these GRBs.
\end{itemize}

The epoch 500s-2000s is shown in panel (b) of Fig. \ref{decays}. During this epoch the GRBs show a slightly higher 
degree of clustering compared with the previous epoch. The hardness ratios of most GRBs transition from highly 
variable to relatively constant during this epoch, with the constant phase indicating that the X-ray and 
optical/UV light curves are produced by a similar mechanism. In  panel (b) of Fig. \ref{decays}, 
5 GRBs are inconsistent with all 5 lines. The rest of the GRBs are consistent with at least one of the 5 lines, implying 
that some GRBs require energy injection. The GRBs are spread evenly above and below the line of equal temporal index, 
indicating that there is no preference for the type of external medium during this epoch, but a single component 
outflow can explain most of the GRBs during this time period. For this epoch, the hardness ratios for most GRBs vary 
more slowly than for the previous epoch and the ratio behaviour in this epoch often continues in to the 2000s-20000s 
epoch. This implies that the period between 500s and 2000s is a transition period where the GRB ceases to have multiple 
emission mechanisms and emission regions and stabilizes to the late time behaviour.

For the epoch 2000s-20000s shown in panel (c), we find that all but three GRBs are consistent with $0<\Delta\alpha\leq0.50$, 
with the majority consistent with $\alpha_{\rm X}+0.25\leq\alpha_{\rm O}\leq\alpha_{\rm X}+0.50$. The consistency 
of most of the GRBs with $\alpha_{\rm X}+0.25\leq\alpha_{\rm O}\leq\alpha_{\rm X}+0.50$ implies that they are satisfied by a constant density medium with 
a cooling break between the X-ray and optical/UV bands. This is also consistent with what was determined using the mean values, 
but the consistency with $0.25<\Delta\alpha\leq0.50$ implies that energy injection is required for these afterglows, although 
$q$ does not appear to have one specific value. The RMS deviations and the hardness ratios indicate that a single synchrotron 
spectrum could produce the optical/UV and X-ray light curves because the X-ray and optical/UV light curves behave in a similar 
way. Four GRBs, are inconsistent with lying below the line of equal temporal index, suggesting that these GRBs lie in a wind medium. 

For the final epoch, $>20000$s, shown in panel (d), the X-ray temporal indices are typically steeper than observed 
for the 2000s-20000s epoch, whereas the range of the optical/UV temporal index has remained the same, implying that for 
at least some GRBs there is a break in the X-ray light curve. Breaks in the X-ray light curves are also seen through 
the tendency of the hardness ratio to slowly decrease. The GRBs in the $>20000$s epoch are mostly consistent with 
$\alpha_{\rm X}+0.25\leq\alpha_{\rm O}\leq\alpha_{\rm X}+0.50$, implying $\nu_{\rm c}$ is between the optical/UV and X-ray bands, the density is 
constant and that energy injection is still required for some GRBs, although possibly fewer than the previous epoch. 
The decreasing hardness ratios indicates a significant difference in the behaviour of the X-ray and optical/UV light curves 
between the 2000s-20000s epoch and the $>20000$s epoch, which could be due to the optical/UV and X-ray lying on separate spectral segments. 
Since in the 2000s-20000s epoch, the GRBs appear to have an arrangement such that $\nu_{\rm m}<\nu_{\rm O}<\nu_{\rm c}<\nu_{\rm X}$, 
it is difficult to produce a decreasing hardness ratio by movement of $\nu_c$, which would move towards either$\nu_{\rm x}$ or $\nu_{\rm o}$. This 
would lead to the X-ray and optical/UV light curves lying on the same spectral segment, which would mean they would have the same 
temporal index and which would lead to a constant hardness ratio rather than a softening one. 
Some GRBs in the 2000s-20000s epoch are consistent with the line of equal temporal index, suggesting that either 
$\nu_{\rm m}<\nu_{\rm O}<\nu_{\rm X}<\nu_{\rm c}$ or $\nu_{\rm m}<\nu_{\rm c}<\nu_{\rm O}<\nu_{\rm X}$, the movement of $\nu_c$ 
between the optical/UV and X-ray would therefore cause a softening or a hardening of the hardness ratio, respectively. For those cases 
where $\nu_{\rm m}<\nu_{\rm O}<\nu_{\rm c}<\nu_{\rm X}$ the decrease in the hardness ratio may be due to differences in the jet 
geometry producing the X-ray and optical/UV component or some form of energy injection may be affecting the relative spectrum. 
Certainly the movement of $\nu_c$ can be excluded since the hardness ratio does not converge to become a constant.
Finally, in the $>20000$s epoch one GRB, GRB~050730, shows evidence of a jet break, with both the X-ray and optical/UV temporal 
indices consistent with the post jet-break temporal relation in Table \ref{Closure_relations}. For GRB~050730, the uncertainties 
on the optical/UV emission are very large, but the hardness ratio decreases slowly, implying that the break may be chromatic, i.e 
occurring only in the X-ray light curve and not the optical/UV light curve.

After 500s there appears to be a cooling break between the optical/UV and X-ray bands for most GRBs and a constant 
density medium is favoured, up to $80\%-90\%$ of the GRBs in panels (c) and (d) of Fig. \ref{decays} are consistent 
with a constant density medium. The favouritism of the X-ray and optical/UV light curves towards a constant density medium is 
also shown by \cite{ryk09}, who compare average decay rates of the X-ray and optical/UV light curves. \cite{cur09b} and 
\cite{pan02}, from samples of 6 \citep[of a total of 10, see][for further details]{cur09b} and 10 well studied GRBs 
respectively, show that approximately half the GRBs are consistent with constant density medium, which is slightly 
lower fraction of GRBs than suggested by this work, at least after 2000s. The higher fraction found in this work and 
\cite{ryk09} may be due to the systematic fitting approach that both works have taken. As for the relative location of the synchrotron cooling frequency with respect 
to the optical/UV and X-ray bands, both \cite{cur09b} and \cite{mel08} independently show that a large fraction of GRBs require a 
spectral break between the optical/UV and X-ray bands, which is typically expected to be $\nu_{O}<\nu_{c}<\nu_{X}$. \cite{cur09b} show that out of 10 GRBs, 
SEDs of eight could be well constrained and 6 of these required a spectral break between the X-ray and optical/UV bands, which 
could be considered to be a cooling break. As for \cite{mel08}, they find that 10 GRBs, from their sample of 24, cannot easily be 
explained by the standard forward shock model. Of the remaining 14 GRBs, 7 appear to have $\nu_{O}<\nu_{c}<\nu_{X}$. The fraction 
of GRBs with $\nu_{O}<\nu_{c}<\nu_{X}$, particularly from \cite{mel08}, is lower than found in this paper, but this paper only 
considers a difference of $0.25\leq\Delta\alpha\leq0.50$ to be due to a cooling frequency and other factors such as 
multi-component jets may contaminate our results. Detailed analysis on a GRB by GRB basis must be used to confirm this result.

While the mean temporal indices form a convincing picture from 500s, an investigation of the individual temporal indices 
in each epoch introduces new aspects to this picture, for instance additional energy injection. The requirement of energy 
injection for some GRBs is also observed through comparison of the spectral and temporal indices of the X-ray light curves 
\citep{eva09}. To complete this picture, we must also look at how the individual GRB light curves change in behaviour 
between the epochs. As observations later than 2000s are expected to probe the emission produced by the jet after it 
has begun to plough into the external medium, which surrounds the progenitor, this emission is less likely to be 
contaminated by emission from the internal shocks. Therefore, we shall examine the change in behaviour between the 
2000s-20000s and $>20000$s epochs.
 
\subsubsection{Implications of the Change in the Temporal Indices Between the 2000s-20000s and $>$20000s epochs}
In Section \ref{XRT/UVOT_temporal_comparison}, we found evidence for chromatic breaks in the afterglows of 7 GRBs. For 
all these GRBs, the breaks occur in the X-ray light curves. Support for this chromatic behaviour can be observed in 
the hardness ratios as a softening, which occurs when the X-ray breaks to a steeper decay, while the optical/UV light 
curve continues to decay at the same rate. The change in the X-ray temporal index and the evolution of the hardness 
ratios provides strong support for chromatic breaks. However, we do caution that a break in the optical/UV light curve at 
late times cannot be excluded without detailed investigation of the afterglows. For each of the 7 GRBs, we fit a power-law 
and a broken power-law to the X-ray light curve from 1000s and onwards. If the broken power-law was the best fit we 
continued to test if a break in the optical/UV light curve could be consistent with the X-ray break. To do this we fit 
a broken power-law to the optical/UV light curve from 1000s onwards, fixing the difference in the temporal index of the two decay segments 
to be the same as found for the X-ray broken power-law fit. We then determined the earliest time at which the optical/UV 
light curve could break and whether this time is consistent with the break in the X-ray light curve. We shifted the break 
time of the fit to the optical light curve so that the $\chi^2$ changed by $\Delta\chi^2=9$ (i.e $3\sigma$). If the 
resulting break time is consistent with the X-ray break time, then the we cannot be certain that the X-ray break is chromatic. Out of 
the 7 GRBs, 5 are best fit by a broken power-law in the X-ray. The 2 other GRBs, GRB~060804 and GRB~060908, could not be fit 
by a broken power-law due to the break occurring before or to close to 1000s. Of the 5 GRBs with X-ray light curves best 
fit by a broken power-law, we are able to convincingly demonstrate that 3 GRBs (GRB~050319, GRB~051109a and GRB~060206) have 
a chromatic break, with the 3$\sigma$ upper limit to an optical break time much later than the X-ray break time.

Achromatic breaks may not truly be achromatic and hence may appear as chromatic breaks. \cite{eer10} have shown through 
simulations that jet breaks, or any variability due to changes in the fluid conditions, may be chromatic, typically 
occurring later in radio bands than in the X-ray or optical. They claim that for certain physical parameters X-ray and 
optical jet breaks (or variability) may occur at different times, although the difference is not well pronounced between 
these two bands. Simulations have also shown that jet breaks may also not be so sharp for lower frequencies compared to 
higher frequencies due to limb brightening effects \citep{granot99,eer10}. This is expected to be most pronounced for 
X-ray/optical versus radio, with the radio emission having the smoothest break. However, the difference in smoothness 
between the X-ray and optical/UV is expected to be less pronounced especially if they lie on the same spectral segments, 
but there may be some difference if $\nu_c$ lies between the two bands. Some achromatic breaks may be confused with 
chromatic breaks due to these effects, however, these effects are likely to cause only minor differences in the break 
times of the optical/UV and X-ray light curves. 

\cite{racusin09} have shown that there is no X-ray spectral evolution after 2000s, therefore breaks which are 
only observed in the X-ray light curve must be due to one of four possibilities: variations in the micro-physical 
parameters \citep{pan06} - which is rather contrived; changes in the external medium - such as was suggested as 
an alternative explanation for GRB~080319B \citep{racusin08}; cessation of energy injection; a jet break. The 
change in the external medium specifically from a constant density to a wind-like medium or vise versa would be 
shown on Fig. \ref{delta} by the GRBs crossing the line of equal temporal index. A position above the line implies 
an ISM-like medium and a position below the line implies a wind-like medium. None of the GRBs with chromatic breaks 
have temporal indices that cross the line of equal temporal index, implying that at least at a simplistic level, the 
change in density of the external medium, from wind-like to constant density or vise versa, can not explain the 
chromatic break. However, this paper has not investigated the relations where $1<p<2$ nor has it investigated 
complex variations in the external density. If we simply apply the closure relations for a constant density medium 
with $\nu_{m}<\nu_{O}<\nu_{c}<\nu_{X}$ to the X-ray and optical/UV temporal indices from the 2000s-20000s and 
$>20000$s epochs for these 7 GRBs then we find for the X-rays $p$ is consistent within $1\sigma$ errors with 
$\geq2$ for 5 GRBs in the 2000s-20000s epoch and $\geq2$ for all 7 GRBs in the $>20000$s epoch. For the 
optical only 3 GRBs are consistent within $1\sigma$ errors with $p\geq2$ in the 2000s-20000s epoch and 3 are 
consistent in the $>20000$s epoch. While this may indicate the $1<p<2$ closure relations should be examined, 
the values of $p$ will increase to $p>2$ if values of $q$, the energy injection parameter, are reduced from 1. 
Since the $1<p<2$ closure relations and changing external media are more complex options they can not be ruled out by 
this work, but shall not be investigated further here. The last two possibilities, cessation of energy injection 
and a jet break, would produce achromatic breaks in a single component outflow. In these cases, changes in temporal 
index of the optical/UV light curves are expected, but these changes are not seen. Therefore, the chromatic breaks 
observed in the X-ray light curves are difficult to explain in terms of a single component outflow. Chromatic 
breaks in several GRBs, which were observed in the X-ray and not the optical/UV (including GRB~050319 and GRB~050802) 
have been investigated by \cite{oates07} and \cite{depas09}, who also found that a single component outflow could 
not explain the observations.

For two GRBs, the temporal indices determined from the epochs 2000s-20000s and $>20000$s lie on different sides of the line of 
equal temporal index, suggesting a change in external density. GRB~061021, crosses from above to below the line of equal temporal 
index, which implies a transition between constant density medium to wind-like medium. Conversely, GRB~070318 crosses from below 
to above the line of equal temporal index, which implies a transition between a wind-like medium to a constant density medium. The 
change in external density essentially changes the frequency of $\nu_c$ \cite[see][for equations describing $\nu_c$ in wind-like 
and constant density media]{zhang04}. For GRB~061021, the X-ray and optical/UV temporal indices, determined from the epochs 
2000s-20000s and $>20000$s, both change by $\geq3\sigma$ and are not consistent with each other. These temporal indices cannot 
be explained by the non-energy injected temporal relations in Table \ref{Closure_relations} with a change in density from constant to wind-like. 
GRB~070318 is also inconsistent with a change in external medium this time from wind-like to constant density because the change 
from wind-like non-energy-injected temporal relations to constant density non-energy-injected temporal relations does not allow 
the X-ray light curves to become steeper while the optical/UV light curves become shallower. Therefore, it is difficult to explain 
why for two GRBs the temporal indices, determined from the epochs 2000s-20000s and $>20000$s, lie on different sides of the line 
of equal temporal index. However, the investigation of external density variations may be too simplistic because the external 
density may have a different density profile and may be highly variable. Temporal relations for $1<p<2$ have also not been examined. 
For $1<p<2$, the temporal indices describing the frequency $\nu_c<\nu$ are different for the constant density and wind-like media. 
This implies that the X-ray and optical/UV temporal indices would always be expected to change, unlike for the $p>2$ case. 
The $1<p<2$ case may be able to explain the behaviour of some of the other GRBs in the sample, especially those that appear to have density changes.

Ultimately, it is difficult to reconcile the optical/UV and X-ray observations of some GRBs in terms of a single 
component jet. We shall now look at more complex geometric models to determine if these can explain 
the observations.

\subsection{Additional Emission Components}
Additional emission components come in two main flavours either the jet consists of two (or more) components or there is 
some form of additional energy injection, such as up-scattered forward shock emission \citep{pan08b} or late `prompt' 
emission \citep{ghi07,ghi09}. 

In a two component jet, there are two theoretical ways in which the optical/UV and X-ray emission can be produced. 
Either the narrow component, with the higher Lorentz factor, produces the X-ray emission and the slower-wider component 
produces the optical/UV emission \citep{oates07,depas09}, or, the narrow and wide components produce both X-ray and optical/UV 
emission \citep{huang04,peng05,granot05}. The simplest scenario is that both components produce X-ray and optical/UV 
emission. However, \cite{oates09} ruled out the possibility because the viewer would observe two peaks from the two 
different emission components. This effect is not seen in the UVOT light curves and therefore, the jet is unlikely 
to have two components where both produce optical/UV emission. 

The second two component jet scenario is that the optical/UV emission is produced by the wide component and the X-ray 
emission is produced by the narrow component. A discussion of how the wide component can produce emission predominantly 
in the optical/UV without contaminating the X-ray and how the narrow component can produce emission predominantly in the 
X-ray without contaminating the optical/UV is provided in \cite{depas09}. In this scenario, the X-ray and optical/UV 
light curves are not required to be produced by the same synchrotron spectrum. However, the X-ray and optical/UV 
afterglows should be satisfied by the temporal relations for the same external medium, either wind-like or ISM-like.

In the up-scattered emission model, the up-scattered emission is thought to be due to photons in the forward shock, 
which travel away from the forward shock towards the outflow. These photons are scattered by interactions with either 
hot or cold electrons in the outflow \citep{pan08b}. If the interactions are with hot electrons then the scattering 
will be Inverse-Compton and seed photons of low energy, will be boosted in to the X-rays. If the interactions are with 
cold electrons then the photons will not gain energy, so a large number of seed photons will be required to be scattered 
to produce sufficient flux to be brighter than the flux of the forward shock. A second effect may cause the up-scattered 
emission to be brighter than the forward shock. If the photons produced at the same time as those in the forward shock 
are up-scattered and received by the observer at a later time after the afterglow has begun to decay, then the scattered 
flux arriving later may be brighter than the forward shock-flux at that time; see \cite{pan08b} for further details. Overall 
the X-ray and optical light curves may be a combination of various degrees of flux contributed from both the forward shock 
and scattering, which enables this model to reproduce flares, plateaus and chromatic breaks. In the case of chromatic 
breaks it would require the scattered emission to cease contributing to the X-ray light curve at the break time, which may 
be difficult to explain. This model has many possibilities for the effect of scattered emission. The scattered emission may 
either not contribute strongly to both the X-ray or optical afterglow, it may contribute strongly to just the X-ray emission, 
or it may contribute strongly to both the X-ray and optical emission. An indication that the scattered emission is dominant 
over the forward shock emission will be a plateau in the obeserved light curves.

In the late `prompt' emission scenario, the central engine is assumed to be active for a period longer than the 
duration of the prompt emission. The central engine steadily produces shells of material at lower and lower Lorentz 
factors, which by internal dissipation produce continuous and smooth emission predominantly in the X-rays, but possibly 
also in the optical/UV \citep{ghi07,ghi09}. The addition of the late `prompt' emission to the afterglow emission 
allows a wide range of temporal indices and allows the model to reproduce a wide range of X-ray and optical/UV temporal 
behaviour including chromatic breaks.

As the late `prompt' emission and the up-scattered emission models predicts light curves that are a combination of two 
different emission components, with varying degrees of contribution from the two components, it is not possible analytically 
to determine if these model are acceptable. However, this wide range in behaviour implies that these scenarios are temporally 
indistinguishable from the two component outflow model. Therefore, in the following we shall talk primarily of whether 
the two component model can explain our observations.

When investigating the single component outflow, we found that the synchrotron cooling frequency typically lies in between 
the X-ray and optical/UV bands, that energy injection may be required for some GRBs, and that there is conclusive evidence 
for a chromatic break in 3 GRBs and evidence for chromatic breaks occurring in 4 further GRBs. These breaks occur in the 
X-ray and cannot easily be explained by a single component outflow. They cannot be explained by a direct change in the external 
density (although complex variation cannot be ruled out), nor by the passage of $\nu_{\rm c}$ 
through the X-ray band because X-ray spectral evolution is not observed during the late afterglow \citep{racusin09}. Therefore, 
as discussed in Section 4.1.3, we consider this break to be due to either the cessation of energy injection or a jet break. 
In a two component outflow, we would expect energy to be injected into both components. However, it is difficult to picture 
the break in the X-rays being caused by the cessation of energy injection in the narrow component only, although from this 
analysis it can not be ruled out completely. Therefore, we take the jet break in the narrow component to be the cause of 
the change in X-ray temporal index \citep{depas09}. However, if this is the case then the X-ray temporal indices after 
$>20000$s are shallower than expected for the uninjected decay post jet-break temporal relations (Table \ref{Closure_relations}). 
The 4th segment of the X-ray light curve, which is considered to be the true post jet-break phase is also shallower than expected 
\citep{eva09}. The inclusion of energy injection will cause the temporal decay index before and after the jet break to be 
less steep. This would be a natural conclusion because energy injection has already been shown to be needed to explain the 
afterglow behaviour of some GRBs. The post jet-break temporal indices from the values predicted in Table \ref{Closure_relations} 
will be reduced by the quantities determined from Eqs. 33, 34 and 35 of \cite{pan06b}. For the simplest jet, a jet with sharp 
edges, which spreads laterally, the temporal index of the post-jet break decay is reduced from $\alpha\sim-p$ by 
$\Delta\alpha=\frac{2}{3}(1-q)(1-\beta)$ for $\nu_{\rm c}<\nu_{\rm X}$. Taking $\beta=-p/2$, then the range in 
$\Delta\alpha$ is $1.33,1.66$ for $q=0$ and $p=2,3$, to $\Delta\alpha=0.0,1.0$ for $q=1$ and $p=2,3$. For $p=2-3$ this 
produces a range $-0.66<\alpha<-3$ for the post-jet break decay. This relation alone can explain the X-ray temporal 
indices of all GRBs for the $>20000$s epoch in Table \ref{GRBs}. The jet may also not show any sideways expansion, in this case the jet is reduced from 
$\alpha=3/2\beta-\frac{2-s}{8-2s}$ by $\Delta\alpha=1/2(1-q)(1-\beta)+\frac{1}{4-s}$ for $\nu_{\rm c}<\nu_{\rm X}$ \citep{pan06b}. 
Again taking $\beta=-p/2$ and $s=0$ indicating a constant denstiy medium, then the range in $\Delta\alpha$ is $1.25,1.50$ for 
$q=0$ and $p=2,3$ to $\Delta\alpha=0.25$ for $q=1$ and $p=2,3$. For $p=2-3$ this produces a range $0<\alpha<-1.75$ for 
the post-jet break decay. This is acceptable for the optical/UV and X-ray temporal indices for the GRBs in the $>20000$s epoch. 
If the post jet break decay is energy injected, then we would expect the 2000s-20000s decay to also be energy injected. In 
this case, the range of temporal indices expected for the 2000s-20000s epoch is given by the energy injected temporal 
relations in Table \ref{Closure_relations} to be $-1.90<\alpha<0.5$, which is consistent with the temporal indices 
determined in this period given in Table \ref{GRBs}.

This appears to be a plausible explanation for the optical/UV and X-ray temporal behaviour of the GRBs with 
chromatic breaks. The wide range of possible temporal indices allowed by the fact the X-ray and optical/UV emission 
are decoupled, implies that the two component model could be used to explain a larger number of GRBs, if not all 
GRBs. However, a comprehensive investigation of the spectral and temporal properties of GRBs is required to determine 
if one of the additional emission mechanisms is able to reproduce all GRB observations.

\section{Conclusions}
\label{conclusions}
In this paper we systematically analyzed a sample of 26 UVOT and XRT observed GRB light curves. We found that the 
behaviour of the optical/UV and X-ray light curves is most different during the early afterglow before 500s, and that 
the light curves behave most similarly during the middle phase of the afterglow between 2000s and 20000s. 

The mean temporal indices of the optical/UV and X-ray light curves determined from three epochs after 500s, imply 
that the average X-ray and optical/UV afterglow is produced by slow cooling electrons, in a constant density 
medium with the synchrotron cooling frequency set between the optical/UV and X-ray bands. However, when we look at 
the individual GRBs, the picture is not so simple. While these properties generally well describe the outflow 
of the individual GRBs from 500s and onwards, this picture requires energy injection to explain the 
temporal indices of some of the GRB outflows. The need for energy injection is shown by the difference 
in the optical/UV and X-ray temporal indices, which require a difference of $0.25\leq\Delta\alpha\leq0.50$, where 
a difference of 0.25 would be expected for non-injected afterglows and 0.50 is the maximum difference expected 
when energy injection is included.

We demonstrated that a chromatic break occurs in the afterglows of three GRBs (GRB~050319, GRB~051109a and GRB~060206), 
while for a further 4 GRB afterglows we have strong indications of chromatic breaks. These breaks are observed in the 
X-ray light curves as a steepening of the X-ray temporal index between 2000s and $10^5$s and a softening of their 
hardness ratios. The lack of X-ray spectral evolution \citep{racusin09} implies these breaks are likely to be caused 
either by changes in the external density, a jet break or is due to the cessation of energy injection. We determined 
that the density evolution on a simplistic scale is not the cause of chromatic breaks, but at this stage we can not rule 
out complex density evolution. Both the jet break and cessation of energy injection would produce an achromatic break if 
the jet is a single component uniform jet. We have shown that chromatic breaks can either be produced if the X-ray and 
optical/UV emission are decoupled and produced in a jet with structure, for instance in a two component jet where the 
narrow component produces the X-ray emission and the wide component produces the optical/UV emission, or it may be 
produced in the late `prompt' emission model or the up-scattered emission model.

\section{Acknowledgments}
This research has made use of data obtained from the High Energy Astrophysics Science Archive Research Center 
(HEASARC) and the Leicester Database and Archive Service (LEDAS), provided by NASA's Goddard Space Flight Center 
and the Department of Physics and Astronomy, Leicester University, UK, respectively. This work made use of data 
supplied by the UK Swift Science Data Centre at the University of Leicester. SRO acknowledges the support of an 
STFC Studentship. MJP, MDP, PAC, NPMK, PAE and KLP acknowledge the support of STFC and SZ thanks STFC for its support through an 
STFC Advanced Fellowship. MMC, TSK, JAN, PWAR and MHS acknowledge support through NASA contract NAS5-00136. We also thank the 
referee for useful comments and suggestions.


\bibliographystyle{mn2e}   
\bibliography{XRT_UVOT} 

\IfFileExists{\jobname.bbl}{}
 {\typeout{}
  \typeout{******************************************}
  \typeout{** Please run "bibtex \jobname" to optain}
  \typeout{** the bibliography and then re-run LaTeX}
  \typeout{** twice to fix the references!}
  \typeout{******************************************}
  \typeout{}
 }

\end{document}